\documentclass[journal,10pt]{IEEEtran}
\IEEEoverridecommandlockouts
\usepackage[nospace,noadjust]{cite}
\usepackage{amsmath,amssymb,amsfonts,times}
\usepackage{graphicx}
\usepackage{textcomp}
\usepackage{xcolor}
\usepackage{subfigure}
\usepackage{epsfig}
\usepackage{multirow}
\usepackage{algorithm}
\usepackage{algorithmicx}
\usepackage{algpseudocode}
\usepackage{array}
\usepackage{epstopdf}
\usepackage{color}
\usepackage{diagbox}
\usepackage{bm}
\usepackage{tcolorbox}
\usepackage{pdfsync}

\usepackage{url}
\usepackage[hidelinks]{hyperref}
\usepackage{amsmath,amsthm,amssymb,amsbsy}
\usepackage{paralist}
\usepackage{xcolor}
\usepackage{color}
\usepackage{graphicx}
\graphicspath{{./figs/}}
\usepackage{comment}
\usepackage{multirow}

\usepackage{graphicx}
\usepackage{fancyhdr}
\usepackage{cite}

\usepackage{cleveref}

\newtheorem{theorem}{Theorem}

\theoremstyle{remark}

\theoremstyle{problem}

\newcommand{\R}{\mathbb{R}}
\def \real    { \mathbb{R} }

\newcommand{\C}{\mathbb{C}}

\newcommand{\e}{\begin{equation}}
\newcommand{\ee}{\end{equation}}
\newcommand{\en}{\begin{equation*}}
\newcommand{\een}{\end{equation*}}
\newcommand{\eqn}{\begin{eqnarray}}
\newcommand{\eeqn}{\end{eqnarray}}
\newcommand{\bmat}{\begin{bmatrix}}
\newcommand{\emat}{\end{bmatrix}}
\renewcommand{\Re}[1]{\operatorname{Re}\left\{#1\right\}}
\renewcommand{\Im}[1]{\operatorname{Im}\left\{#1\right\}}

\DeclareMathAlphabet\mathbfcal{OMS}{cmsy}{b}{n}
\renewcommand{\P}[1]{\operatorname{\mathbb{P}}\left(#1\right)}
\newcommand{\E}{\operatorname{\mathbb{E}}}

\newcommand{\vct}[1]{\boldsymbol{#1}}
\newcommand{\mtx}[1]{\boldsymbol{#1}}

\newcommand{\<}{\langle}
\renewcommand{\>}{\rangle}

\def \vec       {\operatorname*{vec}}
\newcommand{\set}[1]{\mathbb{#1}}

\DeclareMathOperator*{\argmin}{\text{arg~min}}
\DeclareMathOperator*{\argmax}{\text{arg~max}}

\newcommand{\wh}{\widehat}

\newcommand{\wt}{\widetilde}
\newcommand{\ol}{\overline}

\newcommand{\calA}{\mathcal{A}}
\newcommand{\calB}{\mathcal{B}}
\newcommand{\calC}{\mathcal{C}}

\newcommand{\calH}{\mathcal{H}}
\newcommand{\calI}{\mathcal{I}}

\newcommand{\calN}{\mathcal{N}}

\newcommand{\calS}{\mathcal{S}}

\newcommand{\calW}{\mathcal{W}}
\newcommand{\calX}{\mathcal{X}}
\newcommand{\calY}{\mathcal{Y}}

\newcommand{\vr}{\vct{r}}
\newcommand{\vs}{\vct{s}}

\newcommand{\vw}{\vct{w}}
\newcommand{\vx}{\vct{x}}
\newcommand{\vy}{\vct{y}}

\newcommand{\mA}{\mtx{A}}
\newcommand{\mB}{\mtx{B}}

\newcommand{\mG}{\mtx{G}}
\newcommand{\mH}{\mtx{H}}

\newcommand{\mS}{\mtx{S}}

\newcommand{\mW}{\mtx{W}}
\newcommand{\mX}{\mtx{X}}
\newcommand{\mY}{\mtx{Y}}

\newcommand{\mSigma}{\mtx{\Sigma}}

\newcommand{\mId}{{\bf I}}

\newcommand{\setB}{\set{B}}

\graphicspath{{./figs/}}

\newlength{\imgwidth}
\newboolean{twoColVersion}
\setboolean{twoColVersion}{false}
\newcommand{\twoCol}[2]{\ifthenelse{\boolean{twoColVersion}} {#1} {#2} }

\makeatletter
\def\@IEEEsectpunct{\ \,}
\def\paragraph{\@startsection{paragraph}{4}{\z@}{1.5ex plus 1.5ex minus 0.5ex}%
{0ex}{\normalfont\normalsize\sffamily\bfseries}}
\makeatother

\title{Optimal Error Analysis of Channel Estimation \\ for IRS-assisted MIMO Systems}

\author{Zhen~Qin and~Zhihui~Zhu,~\IEEEmembership{Member,~IEEE}

\thanks{
This work was supported by NSF Grants CCF-2241298 and ECCS-2409701. We thank the Ohio Supercomputer Center for providing the computational resources needed in carrying out this work.
(Corresponding author: Zhen Qin and Zhihui Zhu.)

Zhen Qin is with the Michigan Institute for Computational Discovery and Engineering, Department of Electrical Engineering and Computer Science, Department of Statistic, University of Michigan, Ann Arbor, MI 48109 USA, and also with the Department of Computer Science and Engineering, the Ohio State University, Columbus OH 43210 USA. (e-mail: zhenqin@umich.edu). Zhihui Zhu is with the Department of Computer Science and Engineering, the Ohio State University, Columbus OH 43210 USA. (e-mail: zhu.3440@osu.edu).}

}
\begin{document}

\maketitle

\begin{abstract}
As intelligent reflecting surface (IRS) has emerged as a new and promising technology capable of configuring the wireless environment favorably, channel estimation for IRS-assisted multiple-input multiple-output (MIMO) systems has garnered extensive attention in recent years. Despite the development of numerous algorithms to address this challenge, a comprehensive theoretical characterization of the optimal recovery error is still lacking. This paper aims to address this gap by providing theoretical guarantees in terms of stable recovery of channel matrices for noisy measurements. We begin by establishing the equivalence between IRS-assisted MIMO systems in the uplink scenario and a compact tensor train (TT)-based tensor-on-tensor (ToT) regression. Building on this equivalence, we then investigate the restricted isometry property (RIP) for complex-valued subgaussian measurements. Our analysis reveals that successful recovery hinges on the relationship between the number of user terminals and the number of time slots during which channel matrices remain invariant. Utilizing the RIP condition, we establish a theoretical upper bound on the recovery error for solutions to the constrained least-squares optimization problem, as well as a minimax lower bound for the considered model. Our analysis demonstrates that the recovery error decreases inversely with the number of time slots, and increases proportionally with the total number of unknown entries in the channel matrices, thereby quantifying the fundamental trade-offs in channel estimation accuracy. In addition, we explore a multi-hop IRS scheme and analyze the corresponding recovery errors. Finally, we have performed numerical experiments to support our theoretical findings.
\end{abstract}


\section{Introduction}
\label{sec: intro}

\IEEEPARstart{I}{n} a typical wireless propagation environment, transmitted signals undergo attenuation and scattering due to absorption, reflection, unexpected interference, diffraction, and refraction phenomena. Multipath propagation is generally acknowledged as a primary limiting factor in the performance of wireless communication systems \cite{singal2010wireless}. While various physical layer techniques, such as advanced modulation/demodulation and precoding/decoding schemes, alongside the utilization of the mmWave/subTerahertz band, have been devised to counteract these adverse effects at the communication endpoints, it is evident that a plateau has been reached in terms of attainable data rates and performance reliability.

Considering the wireless environment as an additional variable for optimization holds the promise of significant performance gains. This promise is realized through the innovative concept of intelligent reflecting surface (IRS) \cite{tang2020wireless,cao2021reconfigurable,basar2019wireless,gong2020toward,liaskos2018new,jung2020performance,huang2019reconfigurable,basar2020reconfigurable}, which can dynamically reshape the wireless propagation environment to exhibit more desirable characteristics. An IRS is essentially a 2D surface composed of numerous tunable units, which can be implemented using cost-effective antennas or metamaterials and controlled in real-time to manipulate communication channels without emitting signals of their own. Recently, IRS-assisted communications have garnered significant attention \cite{di2020smart}, owing to their potential to enhance efficiency, communication range, and capacity in wireless communication systems. Nevertheless, to fulfill the role of an IRS and achieve efficient and dependable wireless communication, the acquisition of precise channel state information using channel estimation techniques is imperative, posing a formidable challenge. One particular hurdle stems from the assumption that an IRS typically comprises passive elements, thereby necessitating the receiver to estimate the cascaded channel based on pilots transmitted by the transmitter through the IRS. Recently, numerous matrix-based methods for channel estimation \cite{ma2020joint,ning2020channel,wei2020parallel,jensen2020optimal,he2019cascaded,cui2019efficient,chen2023channel, mirza2021channel,jeong2021low,zhao2021exploiting,liu2020matrix,jin2021channel,hu2021two,hu2021semi,wang2020channel} have been proposed and demonstrated commendable performance in practical implementation.

Building on these advancements, another promising approach leverages tensor models for channel estimation. Various tensor-based wireless communication systems, including mmWave \cite{zhou2016channel,wu2020tensor,zhou2017low,lin2020tensor} and IRS-assisted \cite{de2020parafac,de2021channel,zhang2022sparsity,gomes2023channel,du2023time}  multiple-input multiple-output (MIMO) models, have been developed, providing a concise framework for channel estimation. These models aim to accommodate diverse factors like spatial, temporal, spectral, coding, and polarization diversities. Compared to matrix-based models, tensor-based models offer greater flexibility in design, allowing for more effective exploitation of the inherent tensor structure, which can lead to improved recovery accuracy. The canonical polyadic (CP) decomposition \cite{Bro97} was initially applied in the IRS-assisted MIMO model \cite{de2020parafac,de2021channel,zhang2022sparsity} for its compact representation. To convert the optimization problem into a CP factorization problem, an orthogonal pilot signal matrix is assumed, and the IRS-assisted MIMO system needs to satisfy a coefficient constraint, ensuring the uniqueness of factor identification. Due to the lack of efficient and quasi-optimal decomposition methods, the alternating least squares method is commonly employed to tackle this challenge. On the other hand, since CP decomposition combined with an identity tensor can be viewed as a special case of the Tucker decomposition \cite{Tucker66}, \cite{gomes2023channel} constructed a Tucker-based IRS-assisted MIMO model and applied a quasi-optimal method--Higher Order Singular Value Decomposition--to achieve efficient channel estimation.

Various optimization algorithms for IRS-assisted MIMO systems have exhibited strong empirical performance. However, the theoretical analysis of stable recovery has largely focused on identifying the required number of time slots by leveraging conditions that ensure the uniqueness of the solution \cite{de2021channel,pan2022overview,gomes2023channel}.
A lower bound on the estimation error based on the Cram\'er--Rao bound (CRB) was established in \cite{de2021channel}. However, the lower bound could be suboptimal since it is obtained by analyzing a convex surrogate of the original nonconvex least-squares problem, which alters the intrinsic structure of the estimation problem and, consequently, fails to capture its underlying degrees of freedom (see \Cref{Comparison between CRB and minimax lower bound} for a detailed discussion).
To the best of our knowledge, a rigorous optimal error analysis for channel estimation in IRS-assisted MIMO systems is still lacking. This raises the following fundamental question:
\vspace{0.5cm}
\begin{tcolorbox}[colback=white,left=1mm,top=1mm,bottom=1mm,right=1mm,boxrule=.3pt]
{\bf Question}: {What is the optimal recovery error for channel estimation in IRS-assisted MIMO systems, and can this limit be achieved by appropriately formulating the underlying estimation problem?}
\end{tcolorbox}

\subsection{Our contribution:}
In this paper, we affirmatively address the main question through error analyses in the muliple-access/multi-user IRS-assisted MIMO system. We begin by formally establishing that the combination of the channel matrices and the IRS structure can be naturally represented in the tensor train (TT) format \cite{Oseledets11}. Based on this equivalence, we reformulate the IRS-assisted MIMO model as a TT-based tensor-on-tensor (ToT) regression model \cite{llosa2022reduced,qin2024computational}. This reformulation simplifies our analysis and naturally leads to a recovery problem under the TT-based ToT regression model, which captures the uplink communication scenario in a multiple-access/multi-user IRS-assisted MIMO system. To advance this analysis, we initially establish the restricted isometry property (RIP) for complex-valued subgaussian measurements, where each pilot signal element is a complex-valued subgaussian random variable. This analysis reveals an optimal size relationship between user terminals (UTs) and time slots.  We then investigate the recovery of channel matrices via constrained least squares minimization from subgaussian measurements, deriving optimal recovery error bounds that are proportional to the degrees of freedom of the underlying channels. Specifically, we establish an upper bound on the estimation error for constrained least squares objective and further derive a minimax lower bound that depends solely on the underlying IRS-assisted MIMO model, without relying on any specific loss function, thereby further highlighting the optimality of the upper bound. Compared to the CRB, the minimax lower bound more accurately reflects the intrinsic nonconvexity of the model, yielding a tight characterization of the recovery error. Moreover, we extend recovery error considerations to include low-rank channel matrices \cite{de2021channel}, sparse matrices \cite{chen2021hybrid} and unknown IRS \cite{li2020joint}. Next, we explore a multi-hop IRSs scheme and analyze corresponding recovery errors, laying the groundwork for future multi-hop IRSs communication design. Finally, simulations affirm the validity of theoretical analyses.

\subsection{Notations} We use calligraphic letters (e.g., $\calY$) to denote tensors,  bold capital letters (e.g., $\mY$) to denote matrices, except for $\mX_i$ which denotes the $i$-th order-$3$ tensor factors in the TT format ($i=2,\dots,N+M-1$),  bold lowercase letters (e.g., $\vy$) to denote vectors, and italic letters (e.g., $y$) to denote scalar quantities.
$a^*$ is the complex conjugate of $a$.
Elements of matrices and tensors are denoted in parentheses, as in Matlab notation. For example, $\calX(s_1, s_2, s_3)$ denotes the element in position
$(s_1, s_2, s_3)$ of the order-3 tensor $\calX$.
The inner product of $\calA\in\R^{d_1\times\dots\times d_N}$ and $\calB\in\R^{d_1\times\dots\times d_N}$ can be denoted as $\<\calA, \calB \> = \sum_{s_1=1}^{d_1}\cdots \sum_{s_N=1}^{d_N} \calA(s_1,\dots,s_N)\calB(s_1,\dots,s_N) $.
$\|\calX\|_F = \sqrt{\<\calX, \calX \>}$ is the Frobenius norm of $\calX$.
Tensor contraction is defined as $\calA\times_{i}^{j}\calB$ of size $d_1\times\cdots \times d_{i-1} \times \times d_{i+1}\times \cdots\times d_{N} \times h_1 \times \cdots \times h_{j-1} \times h_{j+1}\times \cdots\times h_{M}$ with $(s_1,\dots,s_{i-1},s_{i+1},\dots,s_N,f_1,\dots,f_{j-1},f_{j+1},\dots,f_M)$-th entry being $ \sum_{k}\calA(s_1,\dots,s_{i-1},k,s_{i+1},\dots,s_N)\calB(f_1, \dots, \\ f_{j-1}, k,f_{j+1},\dots,f_M)$ for $\calA\in\R^{d_1 \times \cdots \times d_{i-1}\times d_k \times d_{i+1}\times \cdots\times d_{N}}$ and $\calB\in\R^{h_1 \times \cdots \times h_{j-1}\times d_k \times h_{j+1}\times \cdots\times h_{M}}$.
The procedure of $\calA\times_{i_1,\dots,i_n}^{j_1,\dots,j_n}\calB$ can be viewed as a sequence of $n$ tensor contractions $\calA\times_{i_k}^{j_k}\calB, k\in[n]$.
The symbol $*$ denotes the Khatri-Rao product, and the operator $\vec(\cdot)$ refers to the vectorization of a matrix. $\|\mX\|$ and $\|\mX\|_0$ respectively represent the spectral norm and $\ell_0$ norm of the matrix $\mX$.
For a positive integer $K$, $[K]$ denotes the set $\{1,\dots, K \}$. For two positive quantities $a,b\in \real$, the inequality $b = O(a)$ means $b\leq c a$ for some universal constant $c$; likewise, $b = \Omega(a)$ indicates that $b\ge ca$ for some universal constant $c$.

\section{Single-Hop IRS-Assisted MIMO System}
\label{sec: S-IRS MIMO}

\subsection{Model formulation} We consider a muliple-access/multi-user MIMO communication system assisted by an IRS, in which the $P$ base stations (BSs) receive the signals transmitted by $U$ UTs via the IRS.
The terminology employed in this paper is based on an uplink communication scenario, where the transmitter is referred to as the UT and the receiver as the BS. Without loss of generality, we assume that each BS and each UT are equipped with the same number of antennas, denoted as $L$ and $M$ respectively.  The IRS is composed of $N$ passive elements, or unit cells, capable of individually adjusting their reflection coefficients (i.e., phase shifts).  Due to unfavorable propagation conditions, BS-UT channels are neglected \cite{de2020parafac}. We also assume quasi-static flat fading channel model, where all the channels remain invariant during $T$ time slots. Assuming a block-fading channel, the received signal model is usually given as follows \cite{he2019cascaded,de2020parafac,de2021channel,zhang2022sparsity}
\begin{eqnarray}
    \label{MIMO IRS model}
    \vy(t) = \mH\text{diag}(\vs(t)) \mG \vx(t) + \vw(t), \  t \in[T],
\end{eqnarray}
where the IRS-BS channel is $\mH = \begin{bmatrix}\mH_1^\top  \cdots  \mH_P^\top\end{bmatrix}^\top\in\C^{LP\times N}$ in which  $\mH_p\in\C^{L\times N},p\in[P]$ is the $p$-th IRS-BS channel, and the UT-IRS channel is $\mG = \begin{bmatrix}\mG_1  \cdots  \mG_U\end{bmatrix}\in\C^{N\times UM}$ in which $\mG_u\in\C^{N\times M},u\in[U]$ is $u$-th the UT-IRS channel. $\vx(t) = \begin{bmatrix}\vx_1^\top(t)  \cdots  \vx_U^\top(t)\end{bmatrix}^\top\in\C^{UM\times 1}$ is the vector containing the transmitted pilot signals at time $t$, where $\vx_u(t)\in\C^{M\times 1}$ is the $u$-th user pilot vector. The phase shift vector at time $t$ is defined as $\vs(t) = \begin{bmatrix}s_1(t)e^{j\phi_1(t)}  \cdots  s_N(t)e^{j\phi_N(t)}\end{bmatrix}^\top\in\C^{N\times 1}$, where $s_n(t)\in\{0,1\}$ and $e^{j\phi_n(t)}\in(0,2\pi]$ represent the on or off state and the phase shift of the $n$-th IRS element at time instant $t$, respectively.  $\vw(t) \sim\mathcal{C}\mathcal{N}({\bm 0},\gamma^2\mId_{LP})$ denotes the complex additive white Gaussian noise (AWGN).

The channel training time $T_s$ is divided into $K$ blocks, where each block has $T$ time slots so that $T_s = KT$ \cite{de2020parafac}. Let us define $\vy(k,t) = \vy((k - 1)T + t)$ as the received signal at the $t$-th time slot of the $k$-th block, where $t\in[T]$ and $k\in[K]$. Similarly, we denote $\vx(k,t)$, $\vs(k,t)$ and $\vw(k,t)$ as the pilot signal, phase shift and noise vectors associated with the $t$-th time slot of the $k$-th block. We adopt the structured time-domain training protocol proposed in \cite{de2020parafac}, wherein the IRS phase shift vector remains fixed throughout the $T$ time slots within each block and changes from block to block. Meanwhile, the pilot signals $\begin{bmatrix}\vx(1) \cdots  \vx(T)\end{bmatrix}$ are identically repeated across all $K$ blocks. Consequently, we have
\begin{eqnarray*}
    \label{x and s k t}
\vs(k) &\!\!\!\!=\!\!\!\!& \vs(k,t), t\in[T],\\
    \vx(t) &\!\!\!\!=\!\!\!\!& \vx(k,t), k\in[K].
\end{eqnarray*}
Under these assumptions, the received signal in \eqref{MIMO IRS model} is given by
\begin{eqnarray*}
    \label{MIMO IRS model 1}
    \vy(k,t)\! =\! \mH\text{diag}(\vs(k)) \mG \vx(t) + \vw(k,t), \  t\in[T] \ \text{and} \  k\in[K].
\end{eqnarray*}
By respectively stacking $\vy(k,t)$, $\vx(t)$ and $\vw(k,t),t\in[T]$ into matrices $\mY(k) = \begin{bmatrix}\vy(k,1)  \cdots  \vy(k,T)\end{bmatrix}\in\C^{LP\times T}$, $\mX = \begin{bmatrix}\vx(1) \cdots  \vx(T)\end{bmatrix}\in\C^{UM\times T}$ and $\mW(k) = [\vw(k,1) \cdots  $ $ \vw(k,T)]\in\C^{LP\times T}$, we have
\begin{eqnarray}
    \label{MIMO IRS model 2}
    \mY(k) = \mH\text{diag}(\vs(k)) \mG \mX + \mW(k).
\end{eqnarray}
The objective is to estimate the channel matrices $\mH$ and $\mG$ from the received signals $\{\mY(k)\}_{k=1}^{K}$ by appropriately designing the pilot signal matrix $\mX$ and the phase shift vectors $\{ \vs(k) \}_{k=1}^{K}$. Notably, this formulation highlights $(i)$ the flexibility in designing $\mX$ to optimize estimation performance, and $(ii)$ the desire to minimize the number of time slots $T$ to enhance efficiency.

Previous studies \cite{de2020parafac, de2021channel, lin2021tensor, wei2020parallel, zhang2022sparsity} have reformulated \eqref{MIMO IRS model 2} as a noisy CP factorization problem. This is achieved either by treating $\mG \mX$ as a single matrix \cite{de2020parafac, de2021channel, lin2021tensor} under the assumption that the pseudoinverse of $\mX$ exists, or by assuming that the pilot signal matrix $\mX$ is orthogonal \cite{wei2020parallel,zhang2022sparsity}, such as employing the discrete Fourier transform (DFT) matrix, allowing \eqref{MIMO IRS model 2} to be rewritten as $\mY(k)\mX^{\rm H} = \mH\text{diag}(\vs(k)) \mG  + \mW(k)\mX^{\rm H}$. However, these assumptions impose constraints on the design of the pilot signal matrix, limiting their applicability to general cases. Moreover, the CP factorization framework used for single-hop IRS-assisted MIMO systems does not readily extend to multi-hop configurations, presenting additional challenges.

\subsection{Compact tensor representation}

To accommodate a more general setting for the pilot signal matrix $\mX$ and facilitate the analysis of the optimal recovery error in channel estimation, we can reinterpret \eqref{MIMO IRS model 2} as a tensor-on-tensor (ToT) regression model \cite{lock2018tensor,raskutti2019convex,llosa2022reduced,qin2024computational} which generalizes the classical matrix/tensor regression framework where the response variable is a vector \cite{CandsTIT11, Zhu18TSP, Ma21TSP, Zhu21TIT,qin2024quantum, qin2024guaranteed, qin2024robust,qin2025scalable}.  Specifically, by aggregating the received data over $K$ blocks, we can define three tensors as $\calY\in\C^{LP\times K\times T}$, $\calS\in\C^{N\times K\times N}$ and $\calW\in\C^{LP\times K\times T}$, where $\calY(:,k,:) = \mY(k)$, $\calS(:,k,:) = \text{diag}(\vs(k))$ and $\calW(:,k,:) = \mW(k)$. Then, the ground-truth channel tensor $\calB^\star \in \C^{LP \times K \times UM}$ can be defined as follows:
\begin{eqnarray}
    \label{ground-truth channel tensor}
    \calB^\star(p,k,m) &\!\!\!\!=\!\!\!\!& \mH(p,:) \calS(:,k,:) \mG(:,m)\nonumber\\
    &\!\!\!\! =\!\!\!\!& \mH(p,:)\text{diag}(\vs(k)) \mG(:,m).
\end{eqnarray}
Compared to directly analyzing \eqref{MIMO IRS model 2}, the tensor formulation enables the modeling of correlations in the signal model \eqref{MIMO IRS model 2} across different phase configurations indexed by $k$. Thus, each element of $\calY$  can be represented as
\begin{eqnarray}
    \label{MIMO IRS model 3}
    \calY(p,k,t) = \sum_{m=1}^{UM}\calB^\star(p,k,m) \mX(m,t) + \calW(p,k,t).
\end{eqnarray}
In this context, $\calY$ and $\mX$ represent the first and second tensors in the ``tensor-on-tensor" regression model\footnote{Without loss of generality, we can regard the matrix $\mX$ as an order-$2$ tensor.}, respectively. It is evident that the pilot signal (measurement operator) $\mX$ exclusively affects UTs, as the received signal $\calY$ is a tensor of size $LP\times K\times T$ and the channel tensor $\calB^\star$ in \eqref{MIMO IRS model 3} is a tensor of size $LP\times K\times UM$, implying the total number of antennas in the UTs $UM$ should be contingent on the number of time slots $T$.

Notice that, unlike the construction of CP decomposition, which requires transforming \eqref{ground-truth channel tensor} as demonstrated in \cite{de2020parafac, de2021channel, lin2021tensor, wei2020parallel, zhang2022sparsity}, \eqref{ground-truth channel tensor} is inherently represented in the standard tensor train (TT) format \cite{Oseledets11} as each element of $\calB^\star$ can be expressed in a matrix product form (a detailed introduction to the TT decomposition is provided in the Appendix~\ref{Canonical form for the TT format}). By applying the tensor contraction operation \cite{cichocki2014tensor}, the TT form of $\calB^\star$ can be expressed as:
\begin{eqnarray}
    \label{compact form of TT decomposition}
    \calB^\star = [\mH,\calS,\mG] = \mH\times_{2}^1 \calS \times_{3}^1 \mG \in\C^{LP\times K\times UM},
\end{eqnarray}
where $(\mH\times_{2}^1 \calS)(p,k,i ) = \sum_{n = 1}^N \mH(p,n)\calS(n,k,i)$ and $(\calC \times_{3}^1 \mG )(p,k, m) = \sum_{n = 1}^N \calC(p,k,n)\mG(n,m)$.

Therefore, we can rewrite \eqref{MIMO IRS model 3} as a TT-based ToT regression model.
\begin{eqnarray}
    \label{MIMO IRS model 4}
    \calY = \calX(\calB^\star) + \calW &\!\!\!\!= \!\!\!\!& \calB^\star\times_{3}^1 \mX + \calW\nonumber\\
     &\!\!\!\!=\!\!\!\!& [\mH,\calS,\mG] \times_{3}^1 \mX + \calW,
\end{eqnarray}
where $(\calB^\star\times_{3}^1 \mX)(p,k,t)= \sum_{j=1}^{UM}\calB^\star(p,k,j)\mX(j,t)$. The linear map $\calX(\calB^\star): \C^{LP\times K\times UM} \rightarrow \C^{LP\times K\times T}$ models the measurement process. Here, we emphasize that the TT-based ToT regression model in \eqref{MIMO IRS model 4} serves solely to simplify the theoretical analysis and preserves all the information contained in the original model \eqref{MIMO IRS model 2}.

We now consider minimizing the following constrained least squares objective to recover the channel tensor $\calB^\star$:
\begin{eqnarray}
    \label{The loss function IRS MIMO model}
    \wh \calB = \argmin_{\calB\in\setB_{N,\calS}} \frac{1}{T}\|\calX(\calB) - \calY\|_F^2,
\end{eqnarray}
where we define a set of order-$3$ TT format tensor as following:
\begin{eqnarray}
    \label{The set of TT}
    \setB_{N,\calS} &\!\!\!\!=\!\!\!\!& \{\calB = [\mH,\calS,\mG]: \mH\in\C^{LP\times N}, \mG\in\C^{N\times UM}\ \nonumber\\
    &\!\!\!\!\!\!\!\!& \text{are unknown and}\ \calS\in\C^{N\times K\times N} \ \text{is known} \}.
\end{eqnarray}
The advantage of the least-squares formulation is that it allows the application of well-established tools from standard matrix/tensor regression models, such as the RIP condition \cite{donoho2006compressed, candes2006robust, candes2008introduction,Rauhut17, qin2024quantum}, as detailed in the following discussion. Subsequently, we provide a rigorous theoretical analysis, deriving error bounds for channel estimation.

\subsection{Optimal error analysis} One advantage of viewing the channel estimation as the regression model is that analogous to the analysis in standard regression models, a larger number of time slots $T$ tends to enhance denoising effectiveness since the error bound of channel estimation in the IRS-assisted MIMO model should heavily rely on $T$. Recently, we analyzed the statistical guarantee of the TT-based ToT regression model in \cite{qin2024computational}. Compared with that model, the IRS-assisted MIMO setting introduces additional structural constraints, as the phase-shift tensor $\calS$ is typically known and only the channel matrices are corrupted by noise. Consequently, a direct extension of the results in \cite{qin2024computational} would lead to non-tight recovery bounds that do not properly scale with the degrees of freedom in the unknown channel matrices. Since the recovery error is primarily determined by the channel matrices, we next establish both the upper and minimax lower bounds for the corresponding channel estimation problem.

To facilitate the recovery of channels $\mH$ and $\mG$ from their linear measurements $\calY$, the pilot signal (regression/measurement operator) need adhere to specific properties. One such desirable property is the RIP, extensively explored and popularized in compressive regression literature~\cite{donoho2006compressed, candes2006robust, candes2008introduction, recht2010guaranteed}, and later extended to structured tensor models~\cite{Rauhut17, qin2024quantum}. From a theoretical perspective, the RIP is typically established by assuming that the measurement operators are drawn randomly from a suitable distribution, with subgaussian measurement ensembles serving as a canonical example.
\begin{theorem}
\label{RIP condition fro the TT regression theorem MIMO}
A linear operator $\calX: \C^{LP\times K\times UM} \rightarrow \C^{LP\times K\times T}$ is said to satisfy the $UM$-RIP with a constant $\delta_{UM}$ if
\begin{eqnarray}
    \label{RIP condition fro the TT regression single}
    (1-\delta_{UM})\|\calB\|_F^2\leq \frac{1}{T}\|\mathcal{X}(\calB)\|_F^2\leq(1+\delta_{UM})\|\calB\|_F^2,
\end{eqnarray}
holds for any channel tensor $\calB\in\setB_{N,\calS}$. We now consider two cases:
\begin{itemize}
\item Suppose each element of the pilot signal matrix $\mX$ is an i.i.d. complex-valued subgaussian random variable \footnote{Note that a complex-valued random variable $X$ is subgaussian if and only if its both real part $\Re{X}$ and imaginary part $\Im{X}$ are real subgaussian random variables.
Here are some classical examples of subgaussian distributions.
\begin{itemize}
\item{(Gaussian)} A standard complex Gaussian random variable $X = \Re{X} + i \Im{X}$ with $\Re{X}$ and $\Im{X}$ being independent and following $ \calN(0,\frac{1}{2})$, is a subgaussian random variable.
\item{(Bernoulli)} A Bernoulli random variable $X$ that takes values $-1$ and $1$ with equal probability is a subgaussian random variable.
\item(Digital modulation signals) The phase shift keying (PSK) \cite{li2019optimal}, quadrature amplitude modulation (QAM) \cite{yuan2005analysis} and  pulse amplitude modulation (PAM) \cite{huang2016widely} exhibit subgaussian random variable properties. Specifically, an i.i.d. sequence of PSK, QAM, or PAM symbols can be generated by first drawing a bitstream as an i.i.d. random process with zero mean and bounded variance, and then mapping each bit independently to the corresponding constellation points. The resulting symbol sequence can thus be treated as an i.i.d. complex-valued subgaussian process with a finite subgaussian norm.
\end{itemize}
}. For a positive constant $\delta_{UM}\in(0,1)$, when the number of time slots satisfies
\begin{equation}
T \ge C \cdot \frac{UM}{\delta_{UM}^2},
\label{eq:mrip TT single}
\end{equation}
with probability $1- e^{-c T}$, $\calX$ satisfies the $UM$-RIP with positive constants $c$ and $C$.
\item Suppose $\frac{1}{\sqrt{T}}\mX$ is the normalized DFT matrix, when
\begin{equation}
T \geq UM
\label{eq:mrip TT single for DFT}
\end{equation}
is satisfied, we have $\frac{1}{T}\|\mathcal{X}(\calB)\|_F^2= \|\calB\|_F^2$ with $\delta_{UM} = 0$.
\end{itemize}
\end{theorem}
The proof is presented in the {Appendix}~\ref{Proof of ToT RIP}. \Cref{RIP condition fro the TT regression theorem MIMO} guarantees the RIP for subgaussian measurement ensembles and DFT signals, with the number of time slots $T$ scaling linearly only in relation to the total number of antennas ($UM$) in the UTs. When RIP holds, then for any two distinct TT format channel tensors $\calB_1,\calB_2\in\setB_{N,\calS}$, noting that $\calB_1 - \calB_2 \in\setB_{2N,\calS}$ is also a TT format channel tensor according to (2) in the Appendix~\ref{Canonical form for the TT format}, we have distinct measurements since
\begin{eqnarray}
    \label{diff of two TT formats RIP}
    \frac{1}{T}\|\calX(\calB_1)- \calX(\calB_2)\|_F^2&\!\!\!\! = \!\!\!\!&\frac{1}{T}\|\calX(\calB_1- \calB_2)\|_F^2\nonumber\\
    &\!\!\!\! \ge \!\!\!\!&(1- \delta_{UM}) \|\calB_1- \calB_2\|_F^2,
\end{eqnarray}
which guarantees the possibility of exact recovery of channel matrices in the channel tensor.

Next, we present a formal analysis of the upper bound for $\|\wh \calB - \calB^\star\|_F$ with $\wh\calB \in\setB_{N,\calS}$, obtained by minimizing \eqref{The loss function IRS MIMO model}, independent of the specific optimization algorithm. Here, we provide a concise summary of the main result.
\begin{theorem} (Upper bound of $\|\wh \calB - \calB^\star\|_F$)
\label{Upper bound of error difference MIMO}
Given a channel tensor $\calB^\star$ in \eqref{The set of TT} where $\mH$ and $\mG$ are full rank, when $\calX$ satisfies the $UM$-RIP  and each element in  $\calW$ follows the complex normal distribution $\calC\calN(0,\gamma^2)$, with probability $1-2e^{-c_1(LPN  + UMN)}$ for a positive constant $c_1$, the solution in \eqref{The loss function IRS MIMO model} satisfies
\begin{eqnarray}
    \label{upper bound of error final_ conclusion}
    \|\wh \calB - \calB^\star\|_F=O\bigg(\frac{\gamma\sqrt{(1+\delta_{UM})(LPN  + UMN)}}{(1-\delta_{UM})\sqrt{T}}\bigg),
\end{eqnarray}
where $\wh \calB$ is the solution to \eqref{The loss function IRS MIMO model}.
\end{theorem}
The detailed analysis is provided in the {Appendix}~\ref{Proof of upper bound for error difference MIMO}.
\Cref{Upper bound of error difference MIMO} guarantees a stable recovery of the ground truth $\calB^\star$ when the number of time slots $T$ is linearly proportionate to $LPN  + UMN$. For subgaussian measurements, combining the condition $T = \Omega(UM/\delta_{UM}^2)$ in \eqref{eq:mrip TT single} with $\|\wh \calB - \calB^\star\|_F\leq \epsilon$, the number of time slots $T$ should satisfy $$T = \max\bigg\{\Omega\bigg(\frac{UM}{\delta_{UM}^2}\bigg), \Omega\bigg(\frac{(1+\delta_{UM})(LPN  + UMN)\gamma^2}{(1-\delta_{UM})^2\epsilon^2} \bigg)  \bigg\}$$ to ensure a small recovery error. On the other hand, when using the normalized DFT matrix, to guarantee $\|\wh \calB - \calB^\star\|_F\leq \epsilon$, the number of time slots $T$ should satisfy $$T \geq \max\bigg\{ UM, c\frac{(LPN  + UMN)\gamma^2}{\epsilon^2}   \bigg\},$$ where $c$ is a positive constant.
We note that a commonly considered IRS-assisted MIMO scenario corresponds to $P=1$, i.e., a single base station, under which all theoretical results in this paper remain valid without any modification of the underlying assumptions. In addition, we emphasize that our result does not require any prior knowledge of the channel matrices $\mG$ and $\mH$ or the IRS configuration $\calS$; it only relies on their dimensions. This generality naturally includes special scenarios such as channel hardening \cite{bereyhi2023channel}, which arises when $N$ is sufficiently large.

Next we consider three special cases: $(i)$ \Cref{Upper bound of error difference MIMO} assumes that the channel matrices $\mH$ and $\mG$ are full rank, but in some scenarios, such as millimeter-wave MIMO systems, the presence of a large number of transmit/receive antennas combined with scattering-poor propagation may lead to low-rank channel matrices $\mH$ and $\mG$ \cite{heath2016overview,he2019cascaded}. When the signal travels between the BS and IRS across $r_1$ clusters, and between the IRS and the UT across $r_2$ clusters, it is reasonable to assume that $\mH$ is a rank-$r_1$ matrix and $\mG$ is a rank-$r_2$ matrix. By leveraging the covering number of low-rank matrices as introduced in \cite[Lemma 3.1]{candes2011tight} and applying it to (12) in {Appendix}~\ref{Proof of upper bound for error difference MIMO}, we can readily extend \eqref{upper bound of error final_ conclusion} to
\begin{eqnarray}
    \label{upper bound of error final_ conclusion low-rank}
    &&\hspace{-1.2cm}\|\wh \calB - \calB^\star\|_F\nonumber\\
    &&\hspace{-1.2cm}= O\bigg(\frac{\gamma\sqrt{(1+\delta_{UM})((LP + N)r_1  + (UM + N)r_2)}}{(1-\delta_{UM})\sqrt{T}}\bigg),
\end{eqnarray}
where $\wh \calB,\calB^\star\in\{\calB = [\mH,\calS,\mG]: \mH\in\C^{LP\times N} \ \text{with} \ \text{rank}(\mH) $ \\ $ =r_1, \mG\in\C^{N\times UM} \ \text{with} \ \text{rank}(\mG) = r_2, \text{are unknown and} $ \\ $ \calS\in\C^{N\times K\times N} \ \text{is known} \}$. It is important to note that a comparison between \eqref{upper bound of error final_ conclusion} and \eqref{upper bound of error final_ conclusion low-rank} reveals that the improvement in recovery error due to low-rankness occurs only when $r_1\leq \frac{LPN}{LP + N}$ and $r_2\leq \frac{UMN}{UM + N}$.

$(ii)$ In addition to low-rank structures, channel matrices in mmWave MIMO systems also exhibit sparsity in the angular domain due to the limited number of scattering paths \cite{chen2021hybrid,zhang2022sparsity}. In this case, $\mH$ and $\mG$ can be decomposed as $\mH = \mH_1 \mSigma_1 \mH_2$ and $\mG = \mG_1 \mSigma_2 \mG_2$, where $\mH_1 \in \C^{LP \times F_1}$, $\mH_2 \in \C^{F_1 \times N}$, $\mG_1 \in \C^{N \times F_2}$, and $\mG_2 \in \C^{F_2 \times UM}$ are known matrices defined in \cite{chen2021hybrid}, and $\mSigma_1 \in \C^{F_1 \times F_1}$ and $\mSigma_2 \in \C^{F_2 \times F_2}$ are respectively unknown diagonal matrices containing only $s_1$ and $s_2$ nonzero entries corresponding to the path gains. When $s_1\leq \frac{F_1}{2}$ and $s_2\leq \frac{F_2}{2}$, the sparse channel structure admits an upper bound on the recovery error, given by:
\begin{eqnarray}
    \label{upper bound of error final_ conclusion sparse}
    &&\hspace{-1.2cm}\|\wh \calB - \calB^\star\|_F\nonumber\\
    &&\hspace{-1.2cm}= O\bigg(\frac{\gamma\sqrt{(1+\delta_{UM})(s_1\log(\frac{F_1 }{s_1})  + s_2\log(\frac{F_2 }{s_2}))}}{(1-\delta_{UM})\sqrt{T}}\bigg),
\end{eqnarray}
where $\wh \calB,\calB^\star\in\{\calB = [\mH_1 \mSigma_1 \mH_2,\calS,\mG_1 \mSigma_2 \mG_2]: \mSigma_1 \in \C^{F_1 \times F_1}, \mSigma_2 \in \C^{F_2 \times F_2}, \|\mSigma_1\|_0= s_1, \|\mSigma_2\|_0= s_2 \  \text{are unknown and} \ \calS\in\C^{N\times K\times N}, \mH_1 \in \C^{LP \times F_1}, \mH_2 \in \C^{F_1 \times N}, \mG_1 \in \C^{N \times F_2} \ \text{and} \ \mG_2 \in \C^{F_2 \times UM} \ \text{are known} \}$. The detailed proof has been provided in the {Appendix}~\ref{Proof of upper bound for IRS sparse}. As shown in \eqref{upper bound of error final_ conclusion sparse}, when the logarithmic factor is ignored, the recovery error scales only with the sparsity levels $s_1$ and $s_2$, which explains the practical effectiveness of exploiting sparsity in mmWave MIMO systems.

$(iii)$ It is crucial to emphasize that the elements of the IRS in outdoor scenarios, being exposed to weather and atmospheric conditions, may encounter unknown blockages and time-dependent fluctuations in their phase and amplitude responses \cite{li2020joint}. Consequently, the IRS tensor $\calS$ deviates from its intended structure, and the assumption of perfect knowledge of all phase shifts at the receiver may not be valid in such cases. To analyze the scenario where $\calS$ is unknown, by introducing the covering number for $\calS$ to the proof in {Appendix}~\ref{Proof of upper bound for error difference MIMO}, we can obtain
\begin{eqnarray*}
    \label{upper bound of error final_ conclusion unkown S}
    \|\wh \calB - \calB^\star\|_F= O\big(\frac{\gamma\sqrt{(1+\delta_{UM})(LPN  + KN + UMN)}}{(1-\delta_{UM})\sqrt{T}}\big),
\end{eqnarray*}
where $\wh \calB, \calB^\star\in\{\calB = [\mH,\calS,\mG]: \mH\in\C^{LP\times N}, \calS\in\C^{N\times K\times N}, \mG\in\C^{N\times UM}\ \text{are unknown} \} $.
Apart from the additional term $KN$, this upper bound closely aligns with \eqref{upper bound of error final_ conclusion}. Hence, we omit redundant discussions and concentrate exclusively on the scenario where $\calS$ is perfectly known.

Our second goal is to establish a minimax lower bound using Gaussian ensemble design. This bound serves as a fundamental tool for understanding how the structural properties of a given model govern its achievable recovery error. Notably, this lower bound depends solely on the structure of the model and is independent of any specific loss function or estimator.  Following the derivation in {Appendix}~\ref{proof of minimax bound for TT MIMO}, we arrive at:
\begin{theorem} (Minimax lower bound of $\|\wh{\calB} - \calB^\star\|_F$)
\label{minimax bound of error difference MIMO}
In the context of the TT-based ToT regression model in \eqref{MIMO IRS model 4}, we assume a channel tensor $\calB^\star$ from \eqref{The set of TT}, where $\mH$ and $\mG$ are full rank and $N\geq C'$ with $C'$ being a universal constant. Assume that the noise $\calW$ in \eqref{MIMO IRS model 4} follows $\calC\calN(0,\gamma^2)$, and that the pilot matrix $\mX$ is chosen as either (i) an i.i.d. $\calC\calN(0,1)$  matrix or (ii) a normalized DFT matrix $\frac{1}{\sqrt{T}}\mX$. Under these two choices, we obtain
\begin{eqnarray}
    \label{minimax bound of error final_ conclusion}
    \inf_{\wh \calB}\sup_{\calB^\star\in\setB_{N,\calS}}\E{\|\wh{\calB} - \calB^\star\|_F} = \Omega\bigg(\sqrt{\frac{LPN  + UMN}{T}}\gamma \bigg).
\end{eqnarray}
\end{theorem}
Note that the constraint $N\geq C'$ is not strictly necessary, primarily simplifying the theoretical analysis. By ignoring the constant term associated with $\delta_{UM}$ in \eqref{upper bound of error final_ conclusion}, we conclude that the upper bound--independent of any specific optimization algorithm--on the recovery error of the constrained least squares  minimization is tight and optimal.

\begin{table*}[!ht]
\renewcommand{\arraystretch}{1.7}
\begin{center}
\caption{Comparison of Theoretical Guarantees: Ours vs. \cite{de2021channel}.}
\label{Performance comparison between two references}
{\begin{tabular}{|c||c|c|c|}\hline  {} & {Condition for $T$} & {Upper Bound on Error} &{Lower Bound on Error}
\\\hline\hline {\cite{de2021channel}} & $K\cdot \min\{T, LP\}\geq N, T\geq UM$ (necessary) &  N/A & $\Omega\big(\sqrt{\frac{LPNUM}{T}}\gamma \big)$
\\\hline {Ours} & $T = \Omega\big(\frac{UM}{\delta_{UM}^2}\big)$ or $T\geq UM$ (sufficient) &  $O\big(\sqrt{\frac{(1+\delta_{UM})(LPN  + UMN)}{(1-\delta_{UM})^2T}}\gamma\big) $ & $\Omega\big(\sqrt{\frac{LPN  + UMN}{T}}\gamma \big)$ \\\hline
\end{tabular}}{}
\end{center}
\end{table*}

\subsection{Comparison with Prior Work \cite{de2021channel}}
\label{Comparison between CRB and minimax lower bound}

Similar to our theoretical result, \cite{de2021channel} provides a detailed analysis of the number of time slots required to recover the channel matrices from \eqref{MIMO IRS model 2}, establishing that $K\cdot \min\{T, LP\}\geq N$ and $T\geq UM$ are necessary conditions. However, their analysis primarily concerns a specific bilinear alternating least squares (BALS) algorithm that minimizes a constrained least squares objective. In particular, the condition $K \cdot \min\{T, LP\}\geq N$ ensures the uniqueness of the solution in the CP decomposition, and $T\geq UM$ guarantees that $\mX$ is left-invertible. In contrast, our results in \eqref{eq:mrip TT single} and \eqref{eq:mrip TT single for DFT}, offer  sufficient conditions for exact recovery under constrained least squares minimization in \eqref{The loss function IRS MIMO model}, independent of any specific algorithm.

Moreover, \cite{de2021channel} derives a specific lower bound for the CRB, which has been widely used as a benchmark for performance evaluation in IRS-assisted MIMO channel estimation \cite{de2021channel,du2023time,han2024crb,fernandes2024joint,fernandes2024tensor,yu2025nlos}. Specifically, by defining $\mS = \begin{bmatrix}\vs(1) \cdots \vs(K) \end{bmatrix}^\top\in\C^{K\times N}$, assuming  $\frac{1}{T}\mX\mX^{\rm H} =\mId_{UM}$, and letting each entry of $\calW$ follow the complex normal distribution $\calC\calN(0,\gamma^2)$, the following bound holds:
\begin{eqnarray}
    \label{CRB lower bound derivation}
    \E\|\wh{\calB} - \calB^\star\|_F &\!\!\!\!=\!\!\!\!& \E\|(\wh{\mG}^\top * \wh{\mH}) \mS^\top - (\mG^\top * \mH) \mS^\top\|_F\nonumber\\
    &\!\!\!\!=\!\!\!\!&\Omega\big(\sqrt{\frac{LPNUM}{T}}\gamma \big),
\end{eqnarray}
where the inequality follows from \cite[eq. (70)]{de2021channel}. We note that \eqref{MIMO IRS model 2} is inherently nonconvex, involving a total of $LPN  + UMN$ degrees of freedom in the unknown channel matrices. In general, the recovery error is expected to scale with the number of degrees of freedom \cite{zhang2021preconditioned,han2022optimal,luo2022tensor,qin2024guaranteed,qin2024quantum,qin2024computational}. However, the bound in \eqref{CRB lower bound derivation} scales only with $LPNUM$, as it is derived from a relaxed formulation that transforms the original nonconvex model in \eqref{MIMO IRS model 2} into a convex one by treating $\mG^\top * \mH \in \C^{LPUM \times N}$ as a single unknown variable.As a result, this bound does not accurately reflect the complexity of the original nonconvex problem and is thus not optimal. In contrast, our work provides the first optimal recovery error bound for channel estimation in IRS-assisted MIMO systems that accounts for the full nonconvex structure of the problem. Recall that the minimax lower bound is given by
    \begin{eqnarray}
    \label{minimax bound of error final_ conclusion another}
    \inf_{\wh \calB}\sup_{\calB^\star\in\setB_{N,\calS}}\E{\|\wh{\calB} - \calB^\star\|_F} = \Omega\bigg(\sqrt{\frac{LPN  + UMN}{T}}\gamma \bigg),
    \end{eqnarray}
    which scales with the total degrees of freedom, $LPN + UMN$, in the unknown channel matrices. A detailed comparison is summarized in \Cref{Performance comparison between two references}.

\section{Generalization to Multi-Hop IRS-Assisted MIMO System}
\label{sec: Multi-IRS MIMO}

The preceding section focuses on the single-hop IRS assisted MIMO system, recognized as a promising technology to address the propagation distance challenge \cite{dai2020reconfigurable,liaskos2018new,huang2020holographic,saad2019vision,letaief2019roadmap}. However, in specific scenarios like urban areas or satellite-to-indoor communications, employing a multi-hop IRSs scheme becomes imperative to surmount severe signal blockage between BSs and UTs for enhanced service coverage \cite{huang2021multi,ardah2022double,xu2023coordinating}. Furthermore, in the context of Terahertz (0.1-10 THz) band communication, considered a promising technology for achieving ultra-high speed and low-latency communications, deploying multiple passive IRSs between BSs and UTs proves effective in overcoming inherent propagation attenuations.

\subsection{Tensor-based model formulation} In this section, we explore the $D$-IRS scheme connecting UTs and BSs, assuming unavailability of channels from UTs to BSs, from UTs to the $i$-th IRS ($1<i\leq D$), from the $j$-th IRS ($1\leq j<D$) to BSs, and between non-adjacent IRSs due to blockage or excessive path loss. In detail, we denote the UT-$1$-th IRS channel as $\mB_0\in\C^{N_1\times UM}$, the $D$-th IRS-BS channel as $\mB_D\in\C^{LP\times N_D}$, the channel from the $d$-th IRS to the $(d+1)$-th IRS as $\mB_d\in\C^{N_{d+1}\times N_{d}}$ for $d\in[D-1]$, and represent the $d$-th IRS as $\calS_d\in\C^{N_d\times K\times N_d}$ for $d\in[D]$. Here, $\calS_d(:,k,:) = \text{diag}(\vs_d(k))\in\C^{N_d\times N_d}$, and the design of $\vs_d(k)$ is same with \eqref{MIMO IRS model 2}. Here, we emphasize that, to simplify our theoretical analysis, the IRSs are arranged in a cascaded structure where only the first IRS interacts directly with the UEs, and only the last IRS connects to the BSs. All intermediate IRSs function purely as passive relays and do not directly interface with any BS or UE. A visual illustration can be found in \cite[Figure 1]{huang2021multi}, where, upon excluding the direct path, the remaining configuration accurately represents our model.
We define each element of the channel tensor $\calB^\star_D\in\C^{LP\times K\times UM}$ as follows:
\begin{eqnarray}
    \label{MIMO multi IRS model 4_1}
    \calB^\star_D(p,k,j) &\!\!\!\!=\!\!\!\!& \mB_D(p,:) \Pi_{d = 1}^{D-1}\calS_{d+1}(:,k,:)\mB_{d}\nonumber\\
    &\!\!\!\!\!\!\!\!&\cdot\calS_{1}(:,k,:)\mB_{0}(:,j).
\end{eqnarray}
Assume that each element of $\calW$ is subject to complex additive white Gaussian noise, the received signal is given by
\begin{eqnarray}
    \label{MIMO multi IRS model 3}
    \wt\calY(p,k,t) \! =\! \sum_{j=1}^{UM}\calB^\star_D(p,k,j) \mX(j,t) \!+\! \calW(p,k,t),
\end{eqnarray}
where we differ from the multi-hop scheme in \cite{huang2021multi} by assuming the absence of a direct BS-UT channel. In addition, we neglect both direct and reflected interference across distinct BS-UT links, effectively treating all paths as isolated. Furthermore, signal propagation and interference between different IRSs are neglected.   These assumptions effectively decouple the communication paths, allowing us to concentrate on the highly nonconvex structure of the model in \eqref{MIMO multi IRS model 4_1}.  To the best of our knowledge, a rigorous theoretical characterization of the recovery error for this model remains absent in the literature.

Note that each element $\calB^\star_D(p,k,j)$ can still be conceptualized as an element of an order-$2D+1$ TT format tensor due to its multiplication form resembling (1) in the Appendix~\ref{Canonical form for the TT format}. Specifically, to establish a connection between the channel tensor $\calB^\star_D$ and the TT format tensor, we can reinterpret $\mB_d\in\C^{N_{d+1}\times N_{d}}$ as an order-$3$ tensor $\wt\mB_d\in\C^{N_{d+1} \times 1 \times N_{d}}$. Accordingly, we define $\wh\calB_D^\star$ as the result of a sequence of tensor contractions\footnote{Here, the tensor contraction operation $\calA_1\times_{d}^{1}\calA_2$ for $\calA_1\in\C^{N_1\times\cdots\times N_D}$ and $\calA_2\in\C^{N_1\times M_2\times M_3}$ results in a new tensor $\calA_3$ of size $N_1\times\cdots \times N_{d-1}\times N_{d+1} \times \cdots\times N_D\times M_2\times M_3$, with  the $(n_1,\dots, n_{d-1},n_{d+1},\dots,n_D,m_2,m_3)$-th element being $\sum_{n_d}\calA_1(n_1,\dots,n_d)\calA_2(n_d,m_2,m_3)$.}: $\mB_D\times_{2}^1 \calS_D \times_{3}^1 \wt\mB_{D-1} \times_4^1 \cdots \times_{2D-1}^1\wt\mB_{1} \times_{2D}^1 \calS_{1}\times_{2D+1}^1 \mB_0$. As a result, $\wh\calB_D^\star$ can be represented as an order-$2D+1$ TT format tensor with dimensions $LP\times K\times 1\times K\times \cdots \times 1\times K \times UM$ with TT ranks\footnote{The definition of TT ranks is provided in the Appendix~\ref{Canonical form for the TT format}.} $(N_D, N_D,N_{D-1},N_{D-1},\dots, N_1,N_1)$. Further insights can be gained, revealing that
\begin{eqnarray}
    \label{MIMO multi IRS model 5}
    &\!\!\!\!\!\!\!\!&\hspace{-0.8cm}\calB^\star_D(p,k,j)\nonumber\\
    &\!\!\!\! \!\!\!\!& \hspace{-0.8cm}= \begin{cases}
\wh\calB_D^\star(p,k_1,1,k_2,1,\dots,k_D, j), &  k_1=\dots = k_D = k, \\
0, &  \text{otherwise},
\end{cases}
\end{eqnarray}
where it becomes evident that $\calB^\star_D$ emerges as a sampled outcome of $\wh\calB_D^\star$. To streamline the notation, we introduce the notation $\calB^\star_D = [\mB_D, \calS_D, \mB_{D-1},\dots, \calS_1,\mB_0]\in\C^{LP\times K\times UM}$. Consequently, we define the TT-based ToT regression model as following:
\begin{eqnarray}
    \label{MIMO multi IRS model 4}
    \wt\calY &\!\!\!\!=\!\!\!\!& \calX(\calB^\star_D) + \calW = \calB^\star_D \times_{3}^1 \mX + \calW \nonumber\\
    &\!\!\!\!=\!\!\!\!& [\mB_D, \calS_D, \mB_{D-1},\dots, \calS_1,\mB_0] \times_{3}^1 \mX + \calW.
\end{eqnarray}
Here, we emphasize that the TT-based ToT regression model is equivalent to \eqref{MIMO multi IRS model 3}, and serves to simplify the subsequent error analysis.

Now, we consider the following constrained least squares objective:
\begin{eqnarray}
    \label{The loss function multi IRS MIMO model}
    \wh \calB_D = \argmin_{\calB_D\in\setB_{N,\{\calS_i\}}^D} \frac{1}{T}\|\calX(\calB_D) - \wt\calY\|_F^2,
\end{eqnarray}
where we define the set of $\setB_{N,\{\calS_i\}}^D$ as follows:
\begin{eqnarray}
    \label{The set of multi TT}
\setB_{N,\{\calS_i\}}^D &\!\!\!\!\!\!=\!\!\!\!\!\!&\{\calB_D = [\mB_D,\calS_D,\dots,\mB_1,\calS_1,\mB_0]\in\C^{LP\times K\times UM}:\nonumber\\
     &\!\!\!\!\!\!\!\!\!\!\!\!&\mB_D\in\C^{LP\times N_D}, \mB_0\in\C^{N_1\times UM}, \mB_d\in\C^{N_{d+1}\times N_{d}},\nonumber\\
    &\!\!\!\!\!\!\!\!\!\!\!\!& d\in[D-1] \  \text{are unknown and} \ \calS_d\in\C^{N_d\times K\times N_d}, \nonumber\\
    &\!\!\!\!\!\!\!\!\!\!\!\!&d\in[D] \ \text{are known} \},
\end{eqnarray}
in which $N = \max_d N_d$. Note that each element of $\calB_D$ follows \eqref{MIMO multi IRS model 4_1}. In the subsequent section, we present a comprehensive error analysis of the channel estimation process.

\subsection{Optimal error analysis}
Before presenting the error analysis, we first ensure that the measurements satisfy the RIP condition. By following a similar derivation as provided in {Appendix}~\ref{Proof of ToT RIP}, the RIP condition for $\calX(\calB_D)$ can be directly obtained.
\begin{theorem}
\label{RIP condition fro the TT regression theorem MIMO multi IRS}
A linear operator $\calX: \C^{LP\times K\times UM} \rightarrow \C^{LP\times K\times T}$ is said to satisfy the $UM$-RIP with a constant $\delta_{UM}$ if
\begin{eqnarray}
    \label{RIP condition fro the TT regression multi IRS}
    (1-\delta_{UM})\|\calB_D\|_F^2 \! \leq \! \frac{1}{T}\|\mathcal{X}(\calB_D)\|_F^2 \! \leq \! (1+\delta_{UM})\|\calB_D\|_F^2,
\end{eqnarray}
holds for any channel tensor $\calB_D\in\setB_{N,\{\calS_i\}}^D$. We now consider two cases:
\begin{itemize}
\item Suppose each element of the pilot signal matrix $\mX$ is an i.i.d. complex-valued subgaussian random variable. For a positive constant $\delta_{UM}\in(0,1)$, when the number of time slots satisfies
\begin{equation}
T \ge C \cdot \frac{UM}{\delta_{UM}^2},
\label{eq:mrip TT multi}
\end{equation}
with probability $1- e^{-c T}$, $\calX$ satisfies the $UM$-RIP with positive constants $c$ and $C$.
\item Suppose $\frac{1}{\sqrt{T}}\mX$ is the normalized DFT matrix, when
\begin{equation}
T \geq UM
\label{eq:mrip TT multi for DFT}
\end{equation}
is satisfied, we have $\frac{1}{T}\|\mathcal{X}(\calB_D)\|_F^2\!=\! \|\calB_D\|_F^2$ with $\delta_{UM} = 0$.
\end{itemize}
\end{theorem}
Similar to \Cref{RIP condition fro the TT regression theorem MIMO}, \Cref{RIP condition fro the TT regression theorem MIMO multi IRS} ensures that the number of time slots is solely contingent upon the total number of antennas (UM) in the UTs. This is attributed to the fact that the pilot signal matrix $\mX$ is exclusively associated with the UT-$1$-th IRS channel $\mB_0$. Then we can further analyze the recovery error $\|\wh \calB_D - \calB^\star_D\|_F$.
\begin{theorem} (Upper bound of $\|\wh \calB_D - \calB^\star_D\|_F$)
\label{Upper bound of error difference multi IRS}
Given a channel tensor $\calB^\star_D$ in \eqref{The set of multi TT} in which $\mB_d, d=0,\dots,D$ are full rank, when $\calX$ satisfies the $UM$-RIP and each element in  $\calW$ follows the complex normal distribution $\calC\calN(0,\gamma^2)$, with probability $1-2e^{-c_1( \sum_{d=0}^{D}N_d N_{d+1})\log D}$ for a positive constant $c_1$, the solution in \eqref{The loss function multi IRS MIMO model} satisfies
\begin{align}
    \label{upper bound of error final_multi IRS conclusion}
    \|\wh \calB_D - \calB^\star_D\|_F= O\bigg(\frac{\gamma\sqrt{(1+\delta_{UM})(\sum_{d=0}^{D}N_d N_{d+1})\log D}}{(1-\delta_{UM})\sqrt{T}}\bigg),
\end{align}
where $N_0 = UM $,\! $N_{D+1} = LP $\! and \! $\wh \calB_D$ is the solution to \eqref{The loss function multi IRS MIMO model}.
\end{theorem}
The proof has been provided in {Appendix}~\ref{Proof of upper bound for multi IRS}. \Cref{Upper bound of error difference multi IRS} provides an upper bound on the recovery error derived from a constrained least squares minimization, which scales with the total degrees of freedom of the unknown variables, i.e., $LP N_D+ UM N_1 + \sum_{i=1}^{D-1}N_d N_{d+1} $. However, it is noteworthy that while increasing the number of IRSs can expand service coverage, the recovery error of channel estimation may also increase when the number of time slots $T$ is limited. Therefore, in practical implementation, a balance should be struck between service coverage and recovery error. A discussion of special cases analogous to the single-hop IRS-assisted MIMO system is presented in {Appendix}~\ref{appendix:error analysis for discussion of structured case}.

Building on the upper bound derived from constrained least squares minimization, we now establish a minimax lower bound to assess its optimality.
\begin{theorem} (Minimax lower bound of $\|\wh{\calB}_D - \calB^\star_D\|_F$)
\label{minimax bound of error difference multi MIMO}
In the context of the TT-based ToT regression model in \eqref{MIMO multi IRS model 4}, we assume a channel tensor $\calB^\star_D$ from \eqref{The set of multi TT}, where $\mB_d, d=0,\dots,D$ are full rank and $\min N_d\geq C'$ with $C'$ being a universal constant. Assume that the noise $\calW$ in \eqref{MIMO IRS model 4} follows $\calC\calN(0,\gamma^2)$, and that the pilot matrix $\mX$ is chosen as either (i) an i.i.d. $\calC\calN(0,1)$  matrix or (ii) a normalized DFT matrix $\frac{1}{\sqrt{T}}\mX$. Under these two choices, we obtain
\begin{align}
    \label{minimax bound of error multi final_ conclusion}
    \inf_{\wh \calB_D}\sup_{\calB^\star_D\in\setB_{N,\{\calS_i\}}^D}\E{\|\wh{\calB}_D - \calB^\star_D\|_F} =  \Omega\bigg(\sqrt{\frac{\sum_{d=0}^{D}N_d N_{d+1}}{T}}\gamma \bigg).
\end{align}
\end{theorem}
Our minimax lower bound for the TT-based ToT regression model in \eqref{MIMO multi IRS model 4} demonstrates that the recovery error upper bound is tight and optimal up to the log term.

\setcounter{figure}{1}
\begin{figure*}[!ht]
\centering
\subfigure[]{
\begin{minipage}[t]{0.3\textwidth}
\centering
\includegraphics[width=5.5cm]{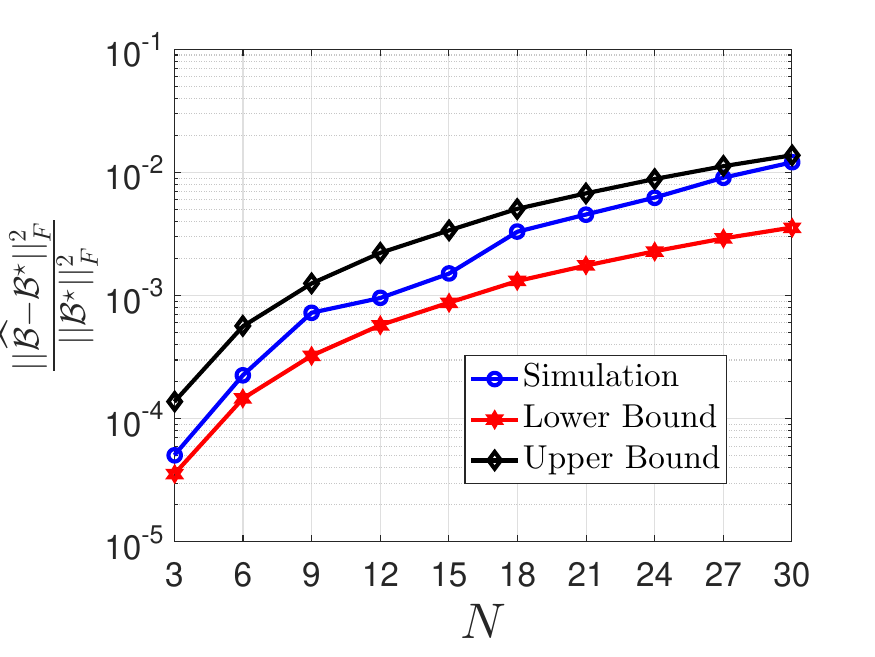}
\end{minipage}
\label{Recovery for N}
}
\subfigure[]{
\begin{minipage}[t]{0.3\textwidth}
\centering
\includegraphics[width=5.5cm]{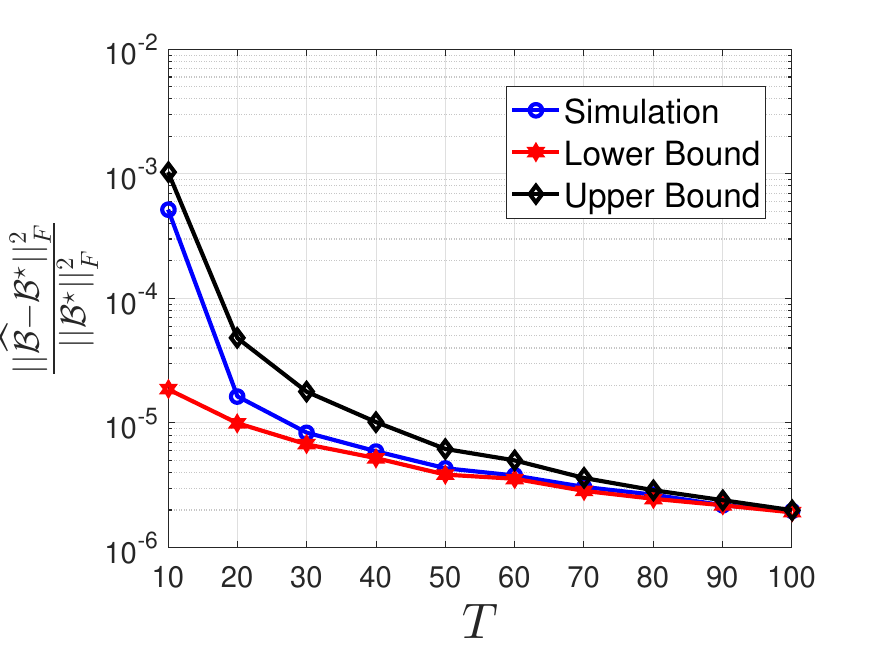}
\end{minipage}
\label{Recovery for T}
}
\subfigure[]{
\begin{minipage}[t]{0.3\textwidth}
\centering
\includegraphics[width=5.5cm]{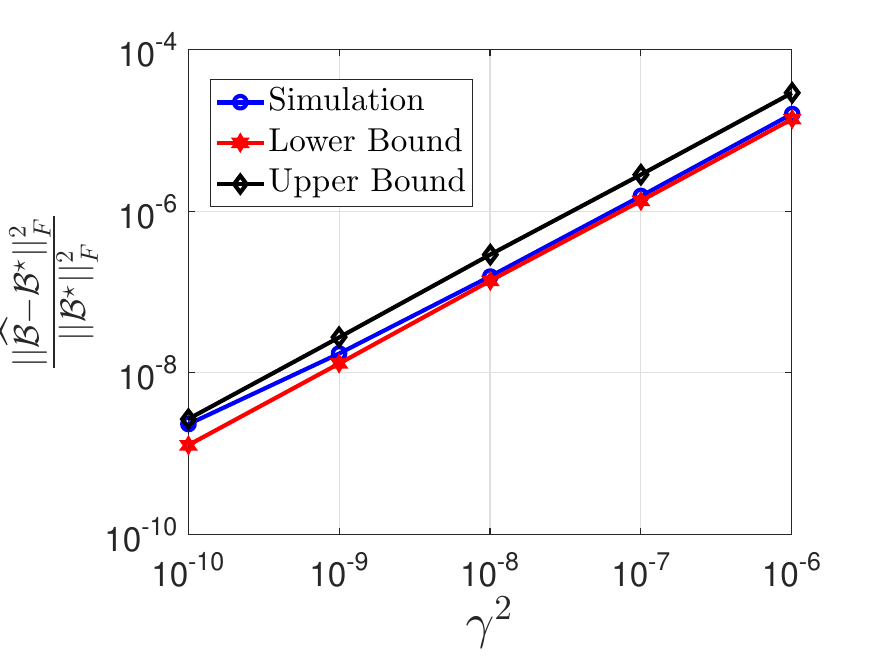}
\end{minipage}
\label{Recovery for noise}
}
\caption{Recovery performance of channel matrices in the single-hop IRS assisted MIMO system (a) for different $N$ with $UM = LP = K  =10 $, $T = 20$ and $\gamma^2 = 10^{-6}$, (b) for different $T$ with  $ UM = LP = K = N =10 $ and $\gamma^2 = 10^{-6}$, (c) for different $\gamma^2$ with  $ UM = LP = K  = N =10$ and $T = 20$.}
\label{Recovery performance for N T noise}
\end{figure*}

\subsection{Discussion} Beyond the IRS-assisted MIMO system, the tensor-based ToT model can also be effectively applied to a variety of typical communication systems, including direct-sequence code-division multiple-access systems \cite{sidiropoulos2000blind}, space-time frequency MIMO systems \cite{favier2014tensor}, and cooperative/relay systems \cite{cavalcante2015joint,favier2016nested}. Compared with traditional representations, tensor-based methods offer a more compact and efficient form for modeling and processing. Furthermore, by following the same analytical approach above, the optimal recovery error for channel estimation in these systems can be derived, as it is proportionate to the degrees of freedom of the channel matrices, assuming the RIP condition is satisfied using certain transmitted pilot signals. These results offer a theoretical guarantee for robustness and potential accuracy in estimating channel characteristics.

\section{Experimental Results}
\label{sec: experiment results}

In this section, we present numerical experiments to validate the effectiveness of the bounds in \Cref{Upper bound of error difference MIMO,minimax bound of error difference MIMO,Upper bound of error difference multi IRS,minimax bound of error difference multi MIMO} for single-hop and multi-hop IRS-assisted MIMO systems\footnote{Here, we replace $\delta_{UM}$ with $c\sqrt{\frac{UM}{T}}$, where $c$ is a positive constant, in accordance with \eqref{eq:mrip TT single} and \eqref{eq:mrip TT multi}. Moreover, since the performance metric in these experiments is the normalized recovery error $\frac{\|\wh\calB - \calB^\star\|_F^2}{\|\calB^\star\|_F^2}$, we accordingly normalize the upper and lower bounds in \Cref{Upper bound of error difference MIMO,minimax bound of error difference MIMO,Upper bound of error difference multi IRS,minimax bound of error difference multi MIMO} by dividing them by $\|\calB^\star\|_F^2$.}. To simplify parameter selection, each entry of the channel matrices and the pilot signal matrix is independently drawn from the standard complex normal distribution. The channel matrices are then normalized to have unit Frobenius norm. Additionally, we set $N_1 = \cdots = N_{D-1} = N$ and denote $\calB_D = \calB$ and $\calB_D^\star = \calB^\star$ for notational simplicity.
To effectively estimate the channel matrices in the single-hop IRS-assisted MIMO system, we adopt the bilinear alternating least squares (BALS) algorithm \cite{de2021channel}, which is designed for constrained least squares minimization and has been widely used as a benchmark in channel estimation \cite{zhang2022sparsity,gomes2023channel,guo2022uplink,de2022semi,de2023semi}. In contrast to BALS, which admits closed-form updates in the single-hop case, the constrained least squares problem in the multi-hop IRS-assisted MIMO system ($D \geq 2$) lacks an analytical solution. To address this, we adopt the alternating gradient descent (AGD) algorithm with learning rate $\mu$, following the gradient-based strategy proposed in \cite{huang2021multi}. The implementation details and computational complexities of both algorithms are provided in the Appendix~\ref{sec: algorithms}. We initialize BALS and AGD with random matrices, with elements generated from a complex standard normal distribution. The number of iterations is set to $100$ for BALS and $10000$ for AGD. For each experimental setting, we perform 20 Monte Carlo trials and report the average results across these trials.

In the first experiment, we investigate the recovery performance of the BALS algorithm under varying values of $UM$ and $LP$, as depicted in \Cref{Performance of different UM and LP figure plot11}. The result illustrates that the recovery error consistently falls within the theoretical range defined by the upper and lower bounds in \Cref{Upper bound of error difference MIMO,minimax bound of error difference MIMO}. In particular, the observed recovery error closely matches the upper bound associated with constrained least squares minimization, which scales proportionally with $LPN + UMN$, as characterized in \Cref{Upper bound of error difference MIMO}. This scaling behavior reflects the intrinsic nonconvex structure of the IRS-assisted MIMO channel estimation problem, where $LPN$ and $UMN$ correspond to the degrees of freedom for the two channel matrices involved in the estimation problem. The empirical agreement with the theoretical bound further confirms the tightness and optimality of our recovery guarantees.

\setcounter{figure}{0}
\begin{figure}[!ht]
\subfigure{
\begin{minipage}[t]{0.44\textwidth}
\centering
\includegraphics[width=6cm]{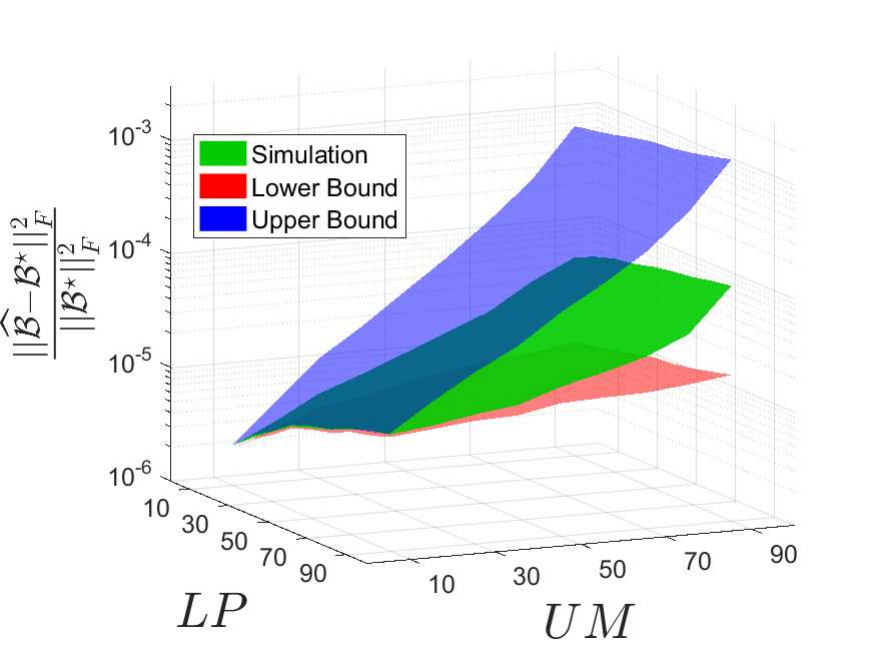}
\end{minipage}
\label{Performance of different UM and LP figure plot11}
}
\caption{Recovery performance of channel matrices in the single-hop IRS assisted MIMO system for different $UM$ and $LP$ with $K = N = 10 $, $T = 100$ and $\gamma^2 = 10^{-6}$.}
\label{Performance of different UM and LP figure plot}
\end{figure}

In the second experiment, we systematically examine the recovery performance of the BALS algorithm across varying values of $N$, $T$ and $\gamma^2$. The empirical findings illustrated in \Cref{Recovery performance for N T noise} demonstrate strong concordance with our theoretical guarantees. Notably, the recovery error exhibits an increasing trend with larger $N$ and $\gamma^2$, and a decreasing trend as the number of time slots $T$ grows, reflecting the fundamental trade-offs inherent in the estimation problem. Crucially, the observed recovery errors are rigorously confined within the analytically derived upper and lower bounds specified in \Cref{Upper bound of error difference MIMO,minimax bound of error difference MIMO}, thereby substantiating the tightness and validity of our theoretical framework.

\setcounter{figure}{2}
\begin{figure}[!ht]
\centering
\includegraphics[width=6cm]{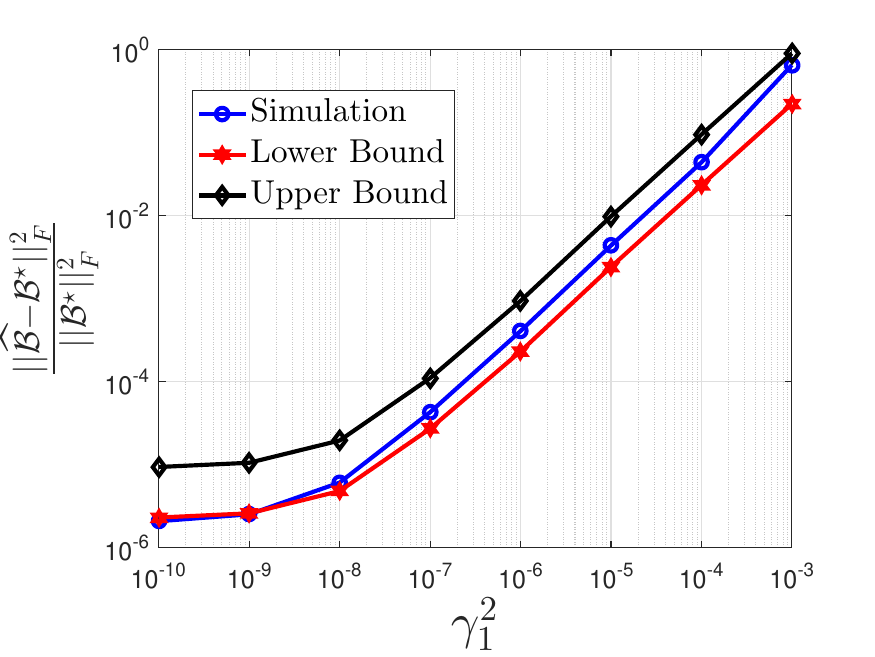}
\caption{Recovery performance of channel matrices in the single-hop IRS assisted MIMO system for different $\gamma_1^2$ with $UM = LP = K = N = 10 $, $\gamma^2 = 10^{-6}$ and $T = 100$.}
\label{Performance of different gamma1}
\end{figure}

In the preceding analysis, the BS-UT direct channels were omitted when evaluating the recovery performance in the single-hop IRS-assisted MIMO system. However, such direct links are often present in practical deployments. To account for this, the model in \eqref{MIMO IRS model 2} can be reformulated as $\mY(k) = \mH\text{diag}(\vs(k)) \mG \mX + \mA \mX +  \mW(k), k\in[K]$, where $\mA \in \C^{LP \times UM}$ represents the BS-UT channel matrix.  In the third experiment, we investigate the impact of the BS-UT channel $\mA$ on the recovery performance of the BALS algorithm. Importantly, the recovery objective remains focused on estimating $\mH$ and $\mG$, rather than jointly estimating $\mH$, $\mG$, and $\mA$. We further assume that each entry of $\mA$ follows $\calC\calN(0,\gamma_1^2)$. As shown in \Cref{Performance of different gamma1}, the recovery error increases as $\gamma_1^2$ becomes larger. This behavior arises because the term $\mA \mX$ effectively acts as an additional noise component when optimizing $\mH$ and $\mG$, similar in effect to $\mW(k)$. Specifically, the $(i,j)$-th entry of $\mA \mX + \mW(k)$ follows a complex Gaussian distribution $\calC\calN(0,\gamma_1^2\|\mX(:,j)\|_2^2 + \gamma^2)$. As a result, the conclusions in \Cref{Upper bound of error difference MIMO,minimax bound of error difference MIMO} can be naturally extended by replacing $\gamma^2$ with $\max_{j}\gamma_1^2\|\mX(:,j)\|_2^2 + \gamma^2$. The empirical recovery errors remain tightly bounded by the theoretically derived upper and lower limits, thus confirming the validity of the theoretical analysis.

In the fourth experiment, we evaluate the recovery performance of the AGD algorithm in multi-hop IRS-assisted MIMO systems with $D=2$, under varying values of $UM$, $LP$, $N$, $T$ and $\gamma^2$. As shown in \Cref{Recovery for UM2,Recovery for LP2,Recovery for N2,Recovery for T2,Recovery for gamma2}, the observed recovery errors exhibit trends that are consistent with the theoretical predictions in \Cref{Upper bound of error difference multi IRS,minimax bound of error difference multi MIMO}. Moreover, the empirical recovery errors remain tightly confined within the theoretically established upper and lower bounds, thereby providing additional evidence for the optimality  of the derived recovery guarantees.

\setcounter{figure}{3}
\begin{figure*}[!ht]
\centering
\subfigure[]{
\begin{minipage}[t]{0.3\textwidth}
\centering
\includegraphics[width=5.5cm]{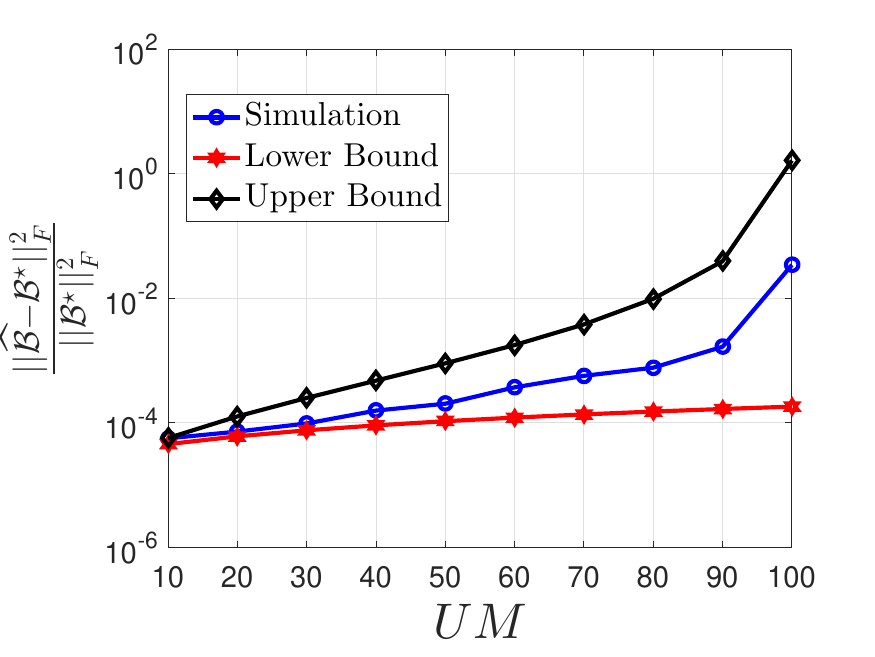}
\end{minipage}
\label{Recovery for UM2}
}
\subfigure[]{
\begin{minipage}[t]{0.3\textwidth}
\centering
\includegraphics[width=5.5cm]{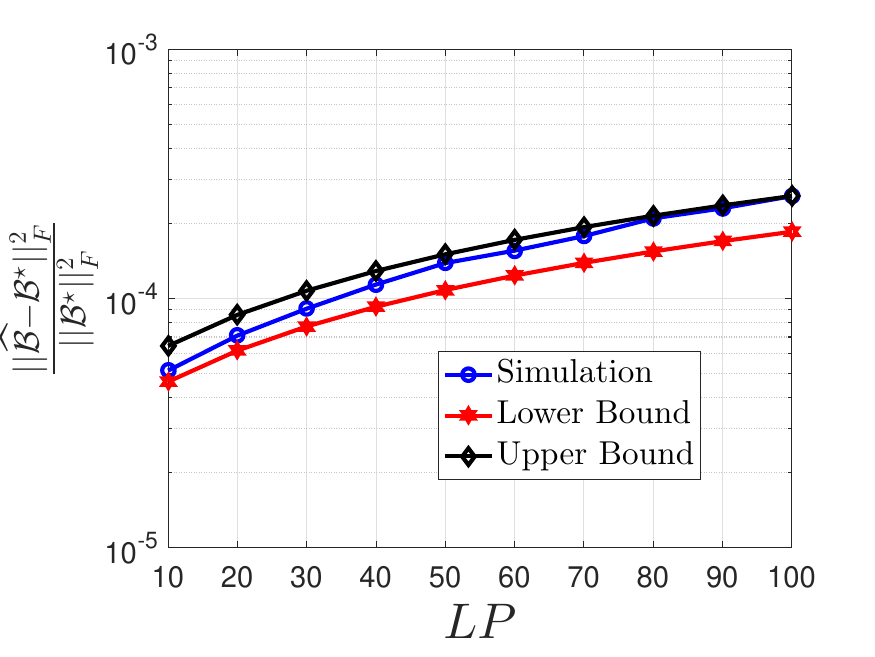}
\end{minipage}
\label{Recovery for LP2}
}
\subfigure[]{
\begin{minipage}[t]{0.3\textwidth}
\centering
\includegraphics[width=5.5cm]{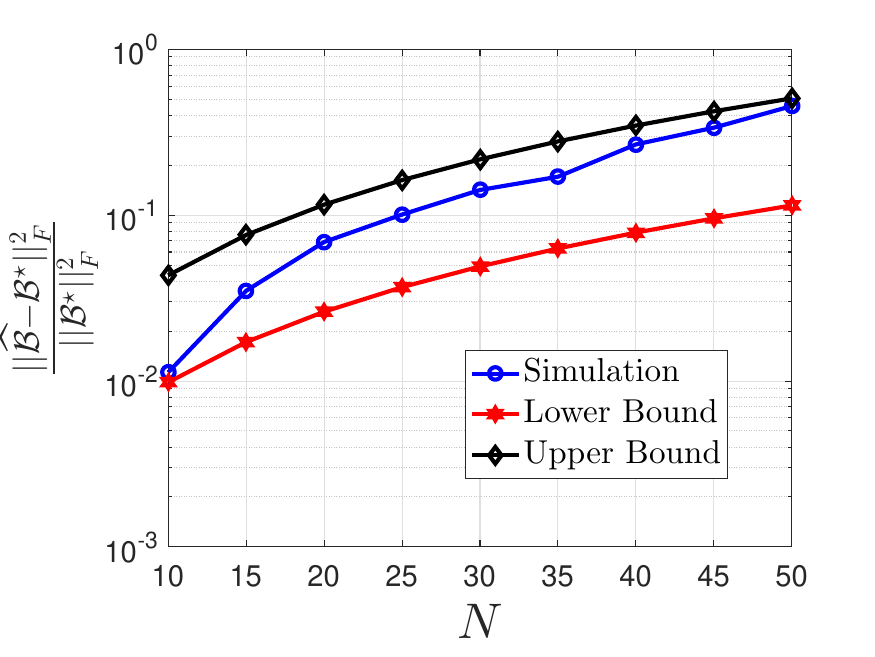}
\end{minipage}
\label{Recovery for N2}
}
\caption{Recovery performance of channel matrices in the multi-hop IRS assisted MIMO system (a) for different $UM$ with $D=2$, $\mu=1$, $LP = K = N =10 $, $T = 100$ and $\gamma^2 = 10^{-6}$, (b) for different $LP$ with $D=2$, $\mu=1$, $UM = K = N =10 $, $T = 100$ and $\gamma^2 = 10^{-6}$, (c) for different $N$ with $D=2$, $\mu=1$, $T = UM =LP =  K =10 $  and $\gamma^2 = 10^{-6}$.}
\label{Recovery performance for T UM LP 2}
\end{figure*}

\setcounter{figure}{4}
\begin{figure*}[!ht]
\centering
\subfigure[]{
\begin{minipage}[t]{0.3\textwidth}
\centering
\includegraphics[width=5.5cm]{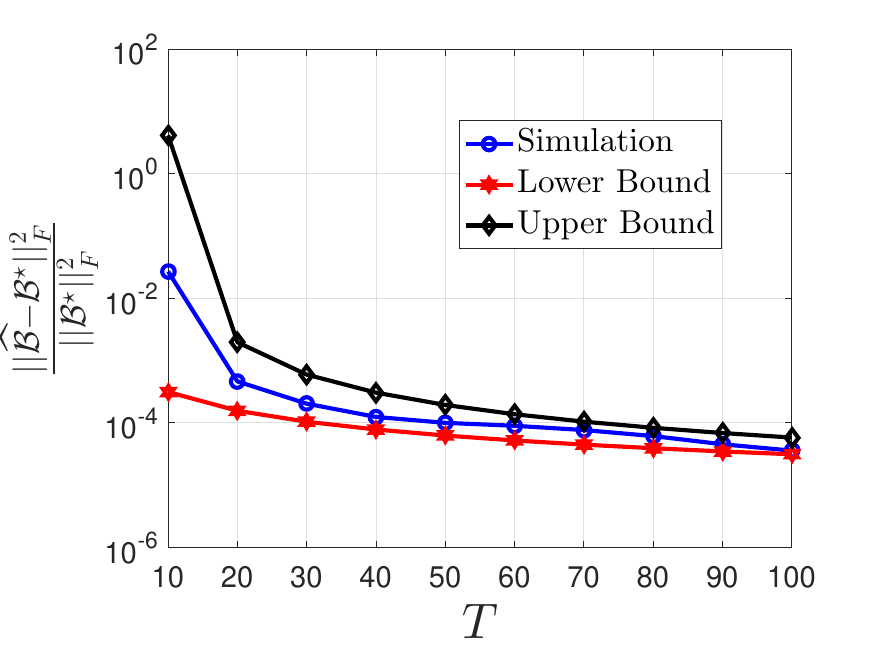}
\end{minipage}
\label{Recovery for T2}
}
\subfigure[]{
\begin{minipage}[t]{0.3\textwidth}
\centering
\includegraphics[width=5.5cm]{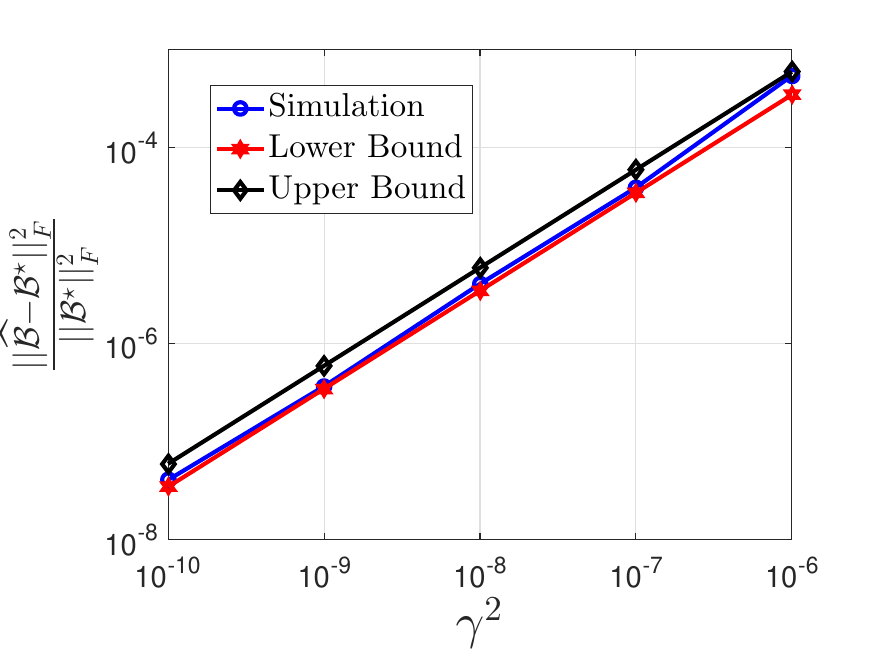}
\end{minipage}
\label{Recovery for gamma2}
}
\subfigure[]{
\begin{minipage}[t]{0.3\textwidth}
\centering
\includegraphics[width=5.5cm]{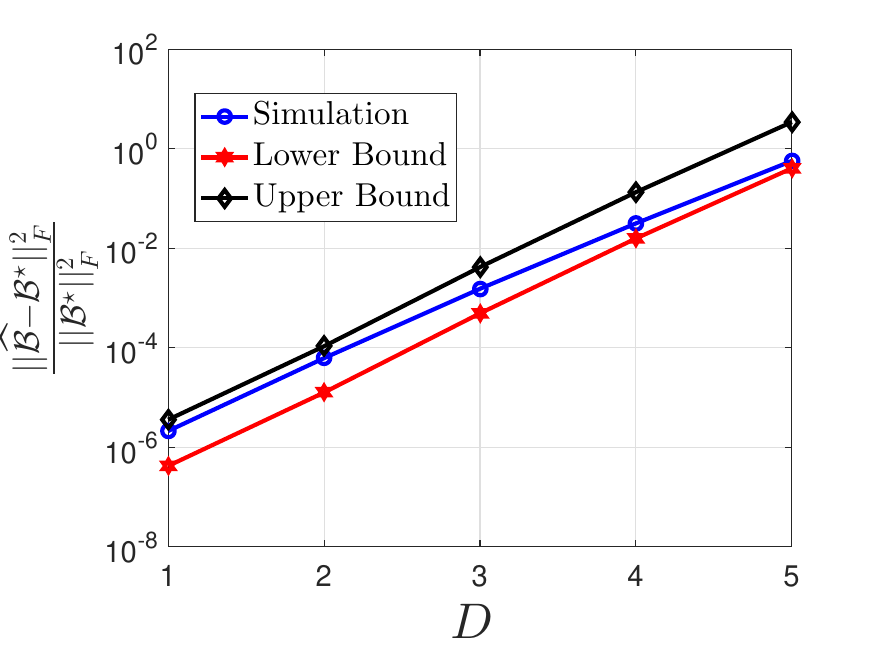}
\end{minipage}
\label{Recovery for D2}
}
\caption{Recovery performance of channel matrices in the multi-hop IRS assisted MIMO system (a) for different $T$ with $D=2$, $\mu=1$, $UM = LP = K = N =10 $ and $\gamma^2 = 10^{-6}$, (b) for different $\gamma^2$ with $D=2$, $\mu=1$, $UM = LP = K = N =10 $ and $T = 20$, (c) for different $D$ with $T = UM =LP =  K  = N =10 $, $\gamma^2 = 10^{-6}$ and $\mu$  sequentially selected from the set $\{1,5,10,15 \}$.}
\label{Recovery performance for T gamma D 2}
\end{figure*}

In the final experiment, we compare the recovery performance of AGD for different numbers of IRSs $D$. As shown in \Cref{Recovery for D2}, the recovery error increases linearly with $D$, aligning with the theoretical analysis in \Cref{Upper bound of error difference multi IRS,minimax bound of error difference multi MIMO}. Moreover, for a fixed number of time slots $T$, the results highlight a fundamental trade-off between enhancing signal propagation through additional IRS layers and maintaining low recovery error. Note that while increasing $D$ can improve end-to-end channel coverage, it simultaneously escalates the difficulty of accurate channel estimation due to the higher dimensionality and the accumulation of noise.

\section{Conclusion}
\label{sec: Conclusion}

This paper addresses the significant gap in optimal error analysis concerning channel estimation for IRS-assisted MIMO systems. By establishing the equivalence between these systems and tensor train-based ToT regression, we provide insights into the fundamental factors crucial for the successful recovery of channel matrices. Specifically, we have established a theoretical upper bound on the recovery error for the constrained least-squares solution, along with a minimax lower bound for the IRS-assisted MIMO model. Our results demonstrate that the recovery error decreases inversely with the number of time slots and increases proportionally with the number of unknown entries in the channel matrices. These findings quantify the fundamental trade-offs inherent in IRS-assisted MIMO channel estimation and provide theoretical justification for the observed empirical behaviors. Additionally, through our exploration of a multi-hop IRS scheme, we evaluate corresponding recovery errors, shedding light on the performance of such configurations. Finally, we validate our theoretical findings through extensive simulations, which consistently align with the derived recovery bounds.

Building upon the results established in this paper, several important research directions can be identified for further investigation.
\begin{itemize}
\item An important area for future research involves conducting error analysis on two-way, two-hop IRS-assisted MIMO systems \cite{zheng2021double,han2022double,le2023double}. Unlike the one-way multi-hop IRS-assisted model discussed in this paper, the optimization problem in the two-way scenario, where channels between base stations/user terminals and IRSs are accessible, consists of two ToT regression problems. Thus, our current analytical approach cannot be directly extended to this setup.

\item Another important challenge is extending our results to the downlink communication scenario. Notably, our recovery guarantees remain applicable in the downlink setting, provided that the RIP holds--achieved by simply interchanging the roles of transmitter and receiver. However, as evidenced by prior studies \cite{choi2014downlink,jiang2015achievable,gu2019information,turan2025versatile}, the number of time slots in the downlink can fall below the number of BSs, thereby violating the RIP condition. This limitation underscores the necessity of developing new analytical frameworks tailored to the downlink regime.

\item Furthermore, recent studies have revealed that even when the direct BS–UT link is blocked, the scattering matrix of the direct channel remains nonzero \cite{nossek2024physically}. Under specific assumptions, the IRS-assisted MIMO model can therefore be reformulated as in \cite[eq. (80)]{nerini2024universal}, which explicitly incorporates the scattering matrix of the reconfigurable impedance network and the IRS reflection (or scattering) coefficient matrix. When these matrices are known, our theoretical results remain directly applicable. However, in practical scenarios, the scattering matrix is typically unknown and the model includes an inverse operation, which prevents the direct application of our current analytical approach. Consequently, extending the error analysis to models without explicit scattering information remains an open and important direction for future work.

\item Finally, a further avenue for exploration is the local convergence analysis and global geometry of the IRS-assisted optimization problem. Despite the promising experimental results obtained even with random initialization, unlike the rotation ambiguity commonly encountered in standard nonconvex matrix/tensor recovery \cite{Zhu18TSP,Ma21TSP,Zhu21TIT,qin2024guaranteed}, a diagonal (scaling) ambiguity exists between channel matrices, as noted in \cite[Section 2]{pan2022overview}. This ambiguity poses a challenge in analyzing the optimization problem at hand.

\end{itemize}

\appendices

\section{Introduction of the TT decomposition}
\label{Canonical form for the TT format}
For an order-$D$ tensor $\calB\in\C^{N_1\times\cdots\times N_D }$, the $(s_1,\dots,s_D)$-th element of $\calB$ in the TT format can be expressed as the following matrix product form \cite{Oseledets11}
\begin{eqnarray}
    \label{Definition of Tensor Train MIMO}
    \calB(s_1,\dots,s_D)=\mB_1(:,s_1,:)\mB_2(:,s_2,:)\cdots \mB_D(:,s_D,:),
\end{eqnarray}
where tensor factors ${\mB}_d \in\C^{r_{d-1}\times N_d \times r_d}, d=1,\dots,D$ with $r_0=r_D=1$.
Thus, the TT format can be represented by $D$ tensor factors $\{{\mB}_d\}_{d\geq 1}$, with a total of $O(D\ol N\ol r^2)$ parameters, where $\ol N = \max_d N_d$ and $\ol r = \max_d r_d$.
In addition, for any two TT format tensors $\wt\calB, \wh\calB\in\C^{N_1\times\cdots\times N_D }$ with factors $\{\wt \mB_d(s_d) \in \C^{\wt r_{d-1}\times \wt r_d} \}$ and $\{ \wh \mB_d(s_d) \in \C^{\wh r_{d-1}\times \wh r_d} \}$, each element of the summation $\calB =  \wt\calB + \wh\calB$ can be represented by
\begin{eqnarray}
    \label{summation of TT format}
    \hspace{0cm}\calB(s_1,\dots,s_D) &\!\!\!\!=\!\!\!\!& \begin{bmatrix}\wt\mB_1(s_1) \!\!\!\! & \wh\mB_1(s_1) \end{bmatrix}
\begin{bmatrix}\wt\mB_2(s_2) \!\!\!\! & {\bm 0} \\ {\bm 0} \!\!\!\! & \wh\mB_2(s_2) \end{bmatrix} \cdots \nonumber\\
&\!\!\!\!\!\!\!\! &\hspace{-1cm}\begin{bmatrix}\wt\mB_{D-1}(s_{D-1}) \!\!\!\! & {\bm 0} \\ {\bm 0} \!\!\!\! & \wh\mB_{D-1}(s_{D-1}) \end{bmatrix} \begin{bmatrix}\wt\mB_D(s_D) \\ \wh\mB_D(s_D) \end{bmatrix},
\end{eqnarray}
which implies that $\calB$ can also be represented in the TT format with ranks  $r_d\leq \wt r_d + \wh r_d$ for $d = 1, \dots, D-1$.

Since any tensor can be decomposed in the TT format \eqref{Definition of Tensor Train MIMO} with sufficiently large TT ranks \cite[Theorem 2.1]{Oseledets11}, the decomposition of a tensor $\calB$ into the form \eqref{Definition of Tensor Train MIMO} is generally not unique: not only are the factors ${\mB_d(:,s_d,:)}$ not unique, but also the dimensions of these factors can vary. To introduce the factorization with the smallest possible dimensions $\vr = (r_1,\ldots,r_{D-1})$, for convenience, for each $d$, we put $\{ {\mB_d(:,s_d,:)}\}_{s_d=1}^{N_d}$ together into the following two forms
\begin{eqnarray}
    \label{left unfolding MIMO}
    &&\hspace{-1cm}L(\mB_d) = \begin{bmatrix}\mB_d(1) \\ \vdots\\  \mB_d(N_d) \end{bmatrix}\in\C^{(r_{d-1}N_d) \times r_d},\\
    \label{right unfolding MIMO}
    &&\hspace{-1cm}R(\mB_d) = \begin{bmatrix}\mB_d(1) &  \cdots &  \mB_d(N_d) \end{bmatrix}\in\C^{r_{d-1}\times (N_dr_d)},
\end{eqnarray}
where $L(\mB_d)$ and $R(\mB_d)$ are often called the left and right unfoldings of $\mB_d$, respectively, if we view $\mB_d$ as a tensor. We say the decomposition \eqref{Definition of Tensor Train MIMO} is minimal if the rank of the left unfolding matrix $L(\mB_d)$ is $r_d$ and the rank of the right unfolding matrix $R(\mB_d)$ is $r_{d-1}$ for all $d$. The dimensions $\vr = (r_1,\dots, r_{D-1})$ of such a minimal decomposition are called the TT ranks of $\calB$. According to \cite{holtz2012manifolds}, there is exactly one set of ranks $\vr$ that $\calB$ admits a minimal TT decomposition.

\section{Nonconvex Iterative Optimization Algorithms}
\label{sec: algorithms}

In this section, we restate the nonconvex iterative algorithms used to estimate the channel matrices within the framework of the TT-based ToT regression model. For a single-hop IRS-assisted MIMO system, we directly adopt the bilinear alternating least squares (BALS) algorithm \cite[Algorithm 2]{de2021channel}. The computational complexity of each iteration of the BALS method, measured in terms of multiplication operations, is given by $O\big(TUMN +  TLPKN + KTN\cdot \min\{KT, N  \}, LPKN\cdot\min\{LPK,N  \} \big)$.
In contrast, for a multi-hop IRS-assisted MIMO system, a closed-form solution for the alternating least squares method is not tractable. Therefore, we primarily consider the alternating gradient descent (AGD) method proposed in \cite{huang2021multi}. Note that we ignore the direct transmission path and focus on minimizing the following loss function:
\begin{eqnarray}
    \label{The loss function multi IRS MIMO model 1}
    &\!\!\!\!\!\!\!\!&\argmin_{\wh\mB_D\in\C^{LP\times N_D},\wh\mB_0\in\C^{N_1\times UM}, \atop \wh\mB_d\in\C^{N_{d+1}\times N_d}, d\in[D-1]}  g(\wh\mB_D,\dots,\wh\mB_0)\nonumber\\
    &\!\!\!\!=\!\!\!\!& \frac{1}{T}\|\calX([\wh\mB_D,\calS_D,\dots,\wh\mB_1,\calS_1,\wh\mB_0]) - \wt\calY\|_F^2.
\end{eqnarray}
We solve this nonconvex optimization problem by the following AGD algorithm for $d=0,\dots,D$:
\begin{eqnarray*}
    \label{GD for factor B i }
    \mB_d^{(l+1)} = \mB_d^{(l)} \! \!- \!\mu\nabla_{\mB_d^*}g(\mB_D^{(l+1)},\dots,\mB_{d+1}^{(l+1)},\mB_d^{(l)},\dots,\mB_0^{(l)}),
\end{eqnarray*}
where $\mu$ is a step size and the Wirtinger gradient $\nabla_{\mB_d^*}g(\mB_D,  \dots,\mB_0)$ is defined as:
\begin{eqnarray}
    \label{gradient for factor B i }
     &\!\!\!\!\!\!\!\!&\hspace{-0.5cm}\nabla_{\mB_d^*}g(\mB_D,\dots,\mB_0) =  \frac{1}{T}\sum_{k=1}^K \calS_{d+1}^{\rm H}(:,k,:)\mB_{d+1}^{\rm H} \cdots \calS_{D}^{\rm H}(:,k,:)\nonumber\\
     &\!\!\!\!\!\!\!\!&\hspace{0.2cm}\cdot\mB_D^{\rm H}\bigg(\calX([\mB_D,\calS_D(:,k,:),\dots,\mB_1,\calS_1(:,k,:), \mB_0])\nonumber\\
    &\!\!\!\!\!\!\!\!&\hspace{0.2cm}- \wt\calY(:,k,:)\bigg)\mX^{\rm H}\mB_{0}^{\rm H}\cdots \mB_{d-1}^{\rm H}\calS_{d}^{\rm H}(:,k,:).
\end{eqnarray}
It is worth noting that AGD in this scenario operates without coefficient constraints, with the only requirement being $T = \Omega(UM)$ to ensure the RIP condition holds. Defining $N = \max_d N_d$, the overall computational complexity of each AGD iteration, measured in terms of multiplication operations, is $O\big(KD(DN^3 + (N + T)LP + UMT(N + LP))\big)$.

\section{Proof of  \Cref{RIP condition fro the TT regression theorem MIMO}}
\label{Proof of ToT RIP}

\begin{proof}
By substituting the concentration inequality from \cite[Theorem 2.3]{candes2011tight} with the concentration inequality for complex-valued subgaussian random variables, when (9) in the main paper holds true, we can assert with a probability of $1-e^{-cT}$ that:
\begin{eqnarray}
    \label{RIP condition fro the TT regression l2}
    (1-\delta_{UM})\|\calB(p,k,:)\|_F^2 &\!\!\!\!\leq\!\!\!\!& \frac{1}{T}\|\mathcal{X}(\calB)(p,k,:)\|_2^2\nonumber\\
    &\!\!\!\!\leq\!\!\!\!&(1+\delta_{UM})\|\calB(p,k,:)\|_F^2.
\end{eqnarray}
Due to $\|\mathcal{X}(\calB)\|_F^2 = \sum_{p,k}\|\mathcal{X}(\calB)(p,k,:)\|_2^2 $ and $\|\calB\|_F^2 = \sum_{p,k}\|\calB(p,k,:)\|_F^2$, this completes the proof for complex-valued subgaussian random variables.

In addition, it is straightforward to see that when $T\geq UM$, the normalized DFT matrix satisfies the RIP.
\end{proof}

\section{Proof of \Cref{Upper bound of error difference MIMO}}
\label{Proof of upper bound for error difference MIMO}

\begin{proof}
Using (7) in the main paper, we have
\begin{eqnarray}
    \label{whX and X star relationship}
    0 &\!\!\!\!\leq\!\!\!\!& \frac{1}{T}\|\calX(\calB^\star) - \calY\|_F^2  - \frac{1}{T}\|\calX(\wh\calB) - \calY\|_F^2\nonumber\\
    &\!\!\!\!=\!\!\!\!& \frac{1}{T}\|\calX(\calB^\star)-\calX(\calB^\star) - \calW\|_F^2\nonumber\\
    &\!\!\!\!\!\!\!\!& - \frac{1}{T}\|\calX(\wh\calB)-\calX(\calB^\star) - \calW\|_F^2\nonumber\\
    &\!\!\!\!=\!\!\!\!& \frac{2}{T}\Re{\<\calX(\calB^\star)+\calW, \calX(\wh\calB - \calB^\star)  \>}\nonumber\\
    &\!\!\!\!\!\!\!\!&+ \frac{1}{T}\|\calX(\calB^\star)\|_F^2 - \frac{1}{T}\|\calX(\wh\calB)\|_F^2\nonumber\\
    &\!\!\!\!=\!\!\!\!& \frac{2}{T}\Re{\<  \calW, \calX(\wh\calB - \calB^\star) \>} - \frac{1}{T}\|\calX(\wh\calB  -  \calB^\star)\|_F^2,
\end{eqnarray}
which further implies that
\begin{eqnarray}
    \label{wX and X star relationship_1}
    \frac{1}{T}\|\calX(\wh\calB  -  \calB^\star)\|_F^2 \leq \frac{2}{T}|\<  \calW, \calX(\wh\calB - \calB^\star) \>|.
\end{eqnarray}

According to (11) in the main paper, we can directly obtain
\begin{eqnarray}
    \label{lower bound of difference}
    \frac{1}{T}\|\calX(\wh\calB  -  \calB^\star)\|_F^2 \geq (1-\delta_{UM})\|\wh\calB  -  \calB^\star\|_F^2.
\end{eqnarray}
For the right-hand side of \eqref{wX and X star relationship_1}, a straightforward application of the Cauchy-Schwarz inequality $|\<  \calW, \calX(\wh\calB - \calB^\star) \>|\leq \|\calW\|_F \|\calX(\wh\calB - \calB^\star) \|_F$ is not adequate to fully elucidate the interplay of all parameters. We address this issue by using the covering argument to bound $|\<  \calW, \calX(\calB - \calB^\star) \>|$ for all possible $\calB$.

Now, we rewrite $\frac{2}{T}|\<  \calW, \calX(\wh\calB - \calB^\star) \>|$ as following:
\begin{eqnarray}
    \label{upper bound of difference}
    &\!\!\!\!\!\!\!\!&\frac{2}{T}|\<  \calW, \calX(\wh\calB - \calB^\star) \>|\nonumber\\
     &\!\!\!\!=\!\!\!\!& \frac{2\|\wh\calB - \calB^\star\|_F}{T}\max_{\calH\in\setB_{2N,\calS}, \|\calH\|_F\leq 1} |\<\calH, \calW\times_3^2 \mX^*\>|\nonumber\\
    &\!\!\!\!=\!\!\!\!&\frac{2\|\wh\calB - \calB^\star\|_F}{T}\hspace{-0.5cm}\max_{\|\mH_1\|\leq 1, \|\mH_2\|_F\leq 1, \atop \|L( \wt\calS)\|\leq 1} \hspace{-0.75cm} |\<[\mH_1,\wt\calS, \mH_2], \calW\times_3^2 \mX^*\>|,
\end{eqnarray}
where $\mX^*$ is the conjugate matrix of $\mX$ and $\calW\times_3^2 \mX^* = \sum_{t=1}^T \calW(:,:,t)\mX^*(:,t)$. According to \cite[eq.(44)]{qin2024guaranteed}, the last line follows $\|\calH\|_F = \|\mH_2\|_F\leq 1$ for a left-orthogonal TT form in {Appendix}~A of the main paper.

To begin, according to \cite{zhang2018tensor}, we can  construct an $\epsilon$-net $\{\mH_1^{(1)} , \dots, \mH_1^{(n_1)}  \}$ with the covering number $n_1\leq (\frac{4+\epsilon}{\epsilon})^{LPN}$ for the set of factors $\{\mH_1\in\C^{LP\times N}:  \|\mH_1\|\leq 1   \}$ such that
\begin{eqnarray}
    \label{H1_proof11}
    \sup_{\mH_1: \|\mH_1\|\leq 1}\min_{p_1\leq n_1} \|\mH_1-\mH_1^{(p_1)}\|\leq \epsilon.
\end{eqnarray}
Similarly, we can construct $\epsilon$-net $\{\mH_2^{(1)} , \dots, \mH_2^{(n_2)}  \}$ with the covering number $n_2\leq (\frac{2+\epsilon}{\epsilon})^{UMN}$ for $\{\mH_2\in\C^{N\times UM}: \|\mH_2\|_F\leq 1  \}$ such that
\begin{eqnarray}
    \label{H3_proof1}
    \sup_{\mH_2: \|\mH_2\|_F\leq 1}\min_{p_{2}\leq n_{2}} \|\mH_2-\mH_2^{(p_{2})})\|_F\leq \epsilon.
\end{eqnarray}
Therefore, we can construct an $\epsilon$-net $\{\calH^{(1)},\ldots,\calH^{(n_1n_2)}\}$ with covering number $n_1n_2 \leq (\frac{4+\epsilon}{\epsilon} )^{LPN  + UMN}$ for any TT format tensor $\calH = [\mH_1,\wt\calS, \mH_{2}]\in\setB_{2N,\wt\calS}$.

Denote by $A$ the value of \eqref{upper bound of difference}, i.e.,
\begin{eqnarray}
    \label{ProofOf<H,X>forSubGaussian_proof3}
    &&\hspace{-1.2cm}[\wt\mH_1,\wt\calS, \wt\mH_2] = \hspace{-0.5cm}  \argmax_{\|\mH_1\|\leq 1,  \atop \|\mH_2\|_F\leq 1,\|L( \wt\calS)\|\leq 1}\hspace{-0.4cm} \frac{1}{T} |\<[\mH_1,\wt\calS, \mH_2], \calW\times_3^2 \mX^* \>|,\\
    \label{ProofOf<H,X>forSubGaussian_proof4}
    &&\hspace{-1.2cm}A:= \frac{1}{T}|\<[\wt\mH_1,\wt\calS, \wt\mH_2], \calW\times_3^2 \mX^*  \>|.
\end{eqnarray}
Using $\calI$ to denote the index set $[n_1]\times [n_2] $, then according to the construction of the $\epsilon$-net, there exists $p=(p_1,p_2)\in\calI$ such that
\begin{eqnarray}
    \label{ProofOf<H,X>forSubGaussian_proof5}
    \|\widetilde\mH_1 - \mH_1^{(p_1)} \|\leq\epsilon, \ \ \  \text{and}  \ \ \  \|\widetilde\mH_2 - \mH_{2}^{(p_{2})}\|_F\leq\epsilon,
\end{eqnarray}
and taking $\epsilon=\frac{1}{4}$ gives
\begin{eqnarray}
    \label{ProofOf<H,X>forSubGaussian_proof6}
    A&\!\!\!\!\leq\!\!\!\!&\frac{1}{T}|\<[\mH_1^{(p_1)},\wt\calS, \mH_2^{(p_2)}], \calW\times_3^2 \mX^*  \>|\nonumber\\
     &\!\!\!\!\!\!\!\!&+ \frac{1}{T}|\<[\wt\mH_1,\wt\calS, \wt\mH_2] - [\mH_1^{(p_1)},\wt\calS, \mH_2^{(p_2)}], \calW\times_3^2 \mX^*  \>| \nonumber\\
    &\!\!\!\!=\!\!\!\!&\frac{1}{T}|\<[\mH_1^{(p_1)},\wt\calS, \mH_2^{(p_2)}], \calW\times_3^2 \mX^*  \>|\nonumber\\
     &\!\!\!\!\!\!\!\!&+ \frac{1}{T}|\< [\wt\mH_1 - \mH_1^{(p_1)}, \wt\calS, \wt\mH_2]\nonumber\\
      &\!\!\!\!\!\!\!\!&+ [\mH_1^{(p_1)}, \wt\calS,\wt\mH_2 - \mH_2^{(p_2)}], \calW\times_3^2 \mX^* \> |\nonumber\\
    &\!\!\!\!\leq\!\!\!\!&\frac{1}{T}|\<[\mH_1^{(p_1)},\wt\calS, \mH_2^{(p_2)}], \calW\times_3^2 \mX^*  \>| + 2 \epsilon A\nonumber\\
    &\!\!\!\!=\!\!\!\!&\frac{1}{T}|\<[\mH_1^{(p_1)},\wt\calS, \mH_2{(p_2)}], \calW\times_3^2 \mX^*  \>| + \frac{A}{2}.
\end{eqnarray}

Note that each element in  $\calW$ follows the complex normal distribution $\calC\calN(0,\gamma^2)$. When conditional on $\mX$, for any fixed $\calH^{(p)} = [\mH_1^{(p_1) }, \wt\calS, \mH_2^{(p_2)}]\in\C^{LP\times K\times UM}$, $\frac{1}{T}\<\calH^{(p)}, \calW\times_3^2 \mX^*  \>$ has complex normal distribution with zero mean and variance $\frac{\gamma^2\|\calX(\calH^{(p)})\|_F^2}{T^2}$, which implies that
\begin{eqnarray}
    \label{the tail function of fixed gaussian random variable}
    \P{\frac{1}{T}|\<\calH^{(p)}, \calW\times_3^2 \mX^*  \>| \geq t | \mX }\leq e^{-\frac{T^2t^2}{2\gamma^2\|\calX(\calH^{(p)})\|_F^2}}.
\end{eqnarray}
Furthermore, under the event $F:=\{\calX \text{ satisfies $UM$-RIP}$ $ \text{  with constant $\delta_{UM}$}\}$, which implies that  $\frac{1}{T}\|\calX(\calH^{(p)})\|_F^2\leq(1+\delta_{UM})\|\calH^{(p)}\|_F^2$. Plugging this together with the fact $\|\calH^{(p)}\|_F\leq 1$ into the above further gives
\begin{eqnarray}
    \label{the tail function of fixed gaussian random variable1}
    \P{\frac{1}{T}|\<\calH^{(p)}, \calW\times_3^2 \mX^*  \>| \geq t | F}\leq e^{-\frac{Tt^2}{2(1+\delta_{UM})\gamma^2}}.
\end{eqnarray}

We now apply this tail bound to \eqref{ProofOf<H,X>forSubGaussian_proof6} and get
\begin{eqnarray}
    \label{the tail function of fixed gaussian random variable 2}
    \P{A \geq t | F} &\!\!\!\!\leq\!\!\!\!& \P{\max_{p_1,p_2} \frac{1}{T}|\<\calH^{(p)}, \calW\times_3^2 \mX^*  \>| \geq \frac{t}{2} | F}\nonumber\\
    &\!\!\!\!\leq\!\!\!\!& \big(\frac{4+\epsilon}{\epsilon}\big)^{2LPN + 2UMN}e^{-\frac{Tt^2}{8(1+\delta_{UM})\gamma^2}}\nonumber\\
    &\!\!\!\!\leq\!\!\!\!& e^{-\frac{Tt^2}{8(1+\delta_{UM})\gamma^2} + c_1(LPN + UMN)},
\end{eqnarray}
where $c_1$ is a constant and based on the assumption $\epsilon=\frac{1}{4}$ in \eqref{ProofOf<H,X>forSubGaussian_proof6}, $\frac{4+\epsilon}{\epsilon}=17$.

Hence, we can take $t = \frac{c_2\sqrt{(1+\delta_{UM})(LPN  + UMN)}}{\sqrt{T}}\gamma$ with a constant $c_2$ and further derive
\begin{eqnarray}
    \label{the tail function of fixed gaussian random variable 3}
    &\!\!\!\!\!\!\!\!&\P{A \leq \frac{c_2\sqrt{(1+\delta_{UM})(LPN  + UMN)}}{\sqrt{T}}\gamma } \nonumber\\
    &\!\!\!\!\geq\!\!\!\!& \P{A \leq \frac{c_2\sqrt{(1+\delta_{UM})(LPN  + UMN)}}{\sqrt{T}}\gamma \cap F } \nonumber\\
    &\!\!\!\!\geq\!\!\!\!&P(F) \P{A \leq \frac{c_2\sqrt{(1+\delta_{UM})(LPN  + UMN)}}{\sqrt{T}}\gamma |F }\nonumber\\
    &\!\!\!\!\geq\!\!\!\!&(1-e^{-c_3UM})(1-e^{-c_4(LPN  + UMN)})\nonumber\\
    &\!\!\!\!\geq\!\!\!\!& 1-2e^{-c_5(LPN  + UMN)},
\end{eqnarray}
where $c_i,i=3,4,5$ are constants. Note that $P(F)$ follows Theorem 1 in the main paper.

Combing \eqref{lower bound of difference}, we can obtain
\begin{eqnarray}
    \label{upper bound of error final}
    \|\wh \calB - \calB^\star\|_F= O\bigg(\frac{\gamma\sqrt{(1+\delta_{UM})(LPN  + UMN)}}{(1-\delta_{UM})\sqrt{T}}\bigg).
\end{eqnarray}
\end{proof}

\section{Proof of \eqref{upper bound of error final_ conclusion sparse} in the main paper}
\label{Proof of upper bound for IRS sparse}
\begin{proof}
Similar to the analysis presented in \Cref{Proof of upper bound for error difference MIMO}, we focus on deriving an upper bound for the term $\frac{2}{T}|\<  \calW, \calX(\wh\calB - \calB^\star) \>|$. To this end, we first rewrite the term as follows:
\begin{eqnarray}
    \label{upper bound of difference sparse}
    \frac{2}{T}|\<  \calW, \calX(\wh\calB - \calB^\star) \>|     &\!\!\!\!=\!\!\!\!& 2/T \hspace{-0.8cm} \max_{\calB\in\{\calC: \calC\in\C^{LP\times K \times UM},\atop \|\calC \|_F\leq \|\wh\calB - \calB^\star\|_F  \} } \hspace{-0.8cm} |\<\calB, \calW\times_3^2 \mX^*\>|.
\end{eqnarray}

Without loss of generality, we assume $\wh \calB,\calB^\star\in \setB =\{\calB = [\mH_1 \mSigma_1 \mH_2,\calS,\mG_1 \mSigma_2 \mG_2]: \mSigma_1 \in \C^{F_1 \times F_1}, \mSigma_2 \in \C^{F_2 \times F_2}, \|\mSigma_1\|_F\leq L_1, \|\mSigma_2\|_F\leq L_2, \|\mSigma_1\|_0= s_1, \|\mSigma_2\|_0= s_2 \  \text{are unknown and} \ \calS\in\C^{N\times K\times N}, \mH_1 \in \C^{LP \times F_1}, \mH_2 \in \C^{F_1 \times N}, \mG_1 \in \C^{N \times F_2} \ \text{and} \ \mG_2 \in \C^{F_2 \times UM} \ \text{are known} \}$.
According to \cite[eq.(2)]{vershynin2009role}, we can construct an $\epsilon_1$-net $\{\mSigma_1^{(1)}, \dots,  \mSigma_1^{(n_1)} \}$ with the covering number $n_1\leq (\frac{CF_1L_1}{s_1\epsilon_1})^{s_1}$ where $C$ is a positive constant, for the set of factors $\{\mSigma_1\in\C^{F_1\times F_1}:  \|\mSigma_1\|_F\leq L_1 \ \text{and} \ \|\mSigma_1\|_0=s_1  \}$ such that
\begin{eqnarray}
    \label{Sigma1 F norm}
    \sup_{\mSigma_1\in\C^{F_1\times F_1}:  \|\mSigma_1\|_F\leq L_1 \atop \ \text{and}\ \|\mSigma_1\|_0=s_1}\min_{p_1\leq n_1} \|\mSigma_1-\mSigma_1^{(p_1)}\|\leq \epsilon_1.
\end{eqnarray}
Similarly, we can construct an $\epsilon_2$-net $\{\mSigma_2^{(1)}, \dots,  \mSigma_2^{(n_2)} \}$ with the covering number $n_2\leq (\frac{CF_2L_2}{s_2\epsilon_2})^{s_2}$   for the set of factors $\{\mSigma_2\in\C^{F_2\times F_2}:  \|\mSigma_2\|_F\leq L_2 \ \text{and} \ \|\mSigma_2\|_0=s_2  \}$ such that
\begin{eqnarray}
    \label{Sigma2 F norm}
    \sup_{\mSigma_2\in\C^{F_2\times F_2}:  \|\mSigma_2\|_F\leq L_2 \atop \ \text{and}\ \|\mSigma_2\|_0=s_2}\min_{p_2\leq n_2} \|\mSigma_2-\mSigma_2^{(p_2)}\|\leq \epsilon_2.
\end{eqnarray}


Therefore, we can construct an $\epsilon$-net $\{\calH^{(1)},\ldots,\calH^{(n_1n_2)}\}$ with covering number $n_1n_2 \!\! \leq \!\! (\frac{CF_1L_1}{s_1\epsilon_1})^{s_1}(\frac{CF_2L_2}{s_2\epsilon_2})^{s_2} $ for any TT format tensor $\calH = [\mH_1 \mSigma_1  \mH_2,\calS,\mG_1  \mSigma_2   \mG_2] $.

Denote by $B$ the value of \eqref{upper bound of difference sparse}, i.e.,
\begin{eqnarray*}
    \label{ProofOf<H,X>forSubGaussian_proof3 sparse}
    &&\hspace{-0.8cm}[\mH_1 \wh\mSigma_1  \mH_2,\calS,\mG_1 \wh\mSigma_2  \mG_2] = \hspace{-0.5cm}  \argmax_{\calB\in\{\calC: \calC\in\C^{LP\times K \times UM},\atop \|\calC \|_F\leq \|\wh\calB - \calB^\star\|_F  \} } \frac{1}{T}|\<\calB, \calW\times_3^2 \mX^*\>|,\\
    \label{ProofOf<H,X>forSubGaussian_proof4 sparse}
    &&\hspace{-0.8cm}B:= \frac{1}{T}|\<[\mH_1 \wh\mSigma_1  \mH_2,\calS,\mG_1 \wh\mSigma_2  \mG_2], \calW\times_3^2 \mX^*\>|.
\end{eqnarray*}

Using $\calI$ to denote the index set $[n_1]\times [n_2] $, then according to the construction of the $\epsilon$-net, there exists $p=(p_1,p_2)\in\calI$ such that
\begin{eqnarray}
    \label{ProofOf<H,X>forSubGaussian_proof5 sparse}
    \|\wh\mSigma_1-\mSigma_1\|\leq \epsilon_1, \ \ \  \text{and}  \ \ \  \|\wh\mSigma_2-\mSigma_2\|\leq \epsilon_2.
\end{eqnarray}
Then taking $\epsilon_1=\frac{L_1}{4}$ and $\epsilon_2=\frac{L_2}{4}$ gives
\begin{eqnarray*}
    B&\!\!\!\!\leq\!\!\!\!&\frac{1}{T}|\<[\mH_1 \mSigma_1^{(p_1)}  \mH_2,\calS,\mG_1 \mSigma_2^{(p_2)}  \mG_2], \calW\times_3^2 \mX^*  \>|\nonumber\\
    &\!\!\!\!\!\!\!\!& + \frac{1}{T}|\<[\mH_1 \wh\mSigma_1  \mH_2,\calS,\mG_1 \wh\mSigma_2  \mG_2]\nonumber\\
    &\!\!\!\!\!\!\!\!&- [\mH_1 \mSigma_1^{(p_1)}  \mH_2,\calS,\mG_1 \mSigma_2^{(p_2)}  \mG_2], \calW\times_3^2 \mX^*  \>| \nonumber\\
    &\!\!\!\!\leq\!\!\!\!&\frac{1}{T}|\<[\mH_1 \mSigma_1^{(p_1)}  \mH_2,\calS,\mG_1 \mSigma_2^{(p_2)}  \mG_2], \calW\times_3^2 \mX^*  \>|\nonumber\\
    &\!\!\!\!\!\!\!\!& + \frac{\epsilon_1}{T}|\<[\mH_1 \frac{(\wh\mSigma_1 - \mSigma_1^{(p_1)})}{\|\wh\mSigma_1 - \mSigma_1^{(p_1)}\|_F}  \mH_2,\calS,\mG_1 \wh\mSigma_2  \mG_2], \calW\times_3^2 \mX^*  \>|\nonumber
\end{eqnarray*}
\begin{eqnarray*}
    &\!\!\!\!\!\!\!\!& + \frac{\epsilon_2}{T}|\<[\mH_1 \mSigma_1^{(p_1)}  \mH_2,\calS,\mG_1 \frac{(\wh\mSigma_2 - \mSigma_2^{(p_2)} )}{\|\wh\mSigma_2 - \mSigma_2^{(p_2)}\|_F} \mG_2], \calW\times_3^2 \mX^*  \>| \nonumber\\
    &\!\!\!\!=\!\!\!\!&\frac{1}{T}|\<[\mH_1 \mSigma_1^{(p_1)}  \mH_2,\calS,\mG_1 \mSigma_2^{(p_2)}  \mG_2], \calW\times_3^2 \mX^*  \>| + \frac{B}{2}.
\end{eqnarray*}

We further assume that $[\mH_1 \mSigma_1^{(p_1)}  \mH_2,\calS,\mG_1 \mSigma_2^{(p_2)}  \mG_2]\in\setB\cap \{ \calB = [\mH_1 \mSigma_1 \mH_2,\calS,\mG_1 \mSigma_2 \mG_2]: \|\calB\|_F\leq\|\wh\calB - \calB\|_F \}$. Following the same analysis of \eqref{upper bound of error final}, with high probability, we have
{\small \begin{eqnarray*}
    \label{upper bound of error final sparse}
    \|\wh \calB - \calB^\star\|_F= O\bigg(\frac{\gamma\sqrt{(1+\delta_{UM})(s_1\log(\frac{F_1}{s_1})  + s_2\log(\frac{F_2}{s_2}))}}{(1-\delta_{UM})\sqrt{T}}\bigg).
\end{eqnarray*}}

\end{proof}

\section{Proof of \Cref{minimax bound of error difference MIMO}}
\label{proof of minimax bound for TT MIMO}

\begin{proof}
Suppose we can find a set of $\{\calB^j  \}_{j=1}^n\in\setB_{N,\calS}$ such that $\min_{j\neq k}\|\calB^j - \calB^k\|_F\geq s$.
According to \cite[Theorem 4]{luo2022tensor}, when each element of $\mX$ and $\calW$ respectively follow $\calC\calN(0,1)$ and $\calC\calN(0,\gamma^2)$, we have
\begin{eqnarray}
    \label{lower bound of the minimax}
    & \!\!\!\!\!\!\!\!&\inf_{\wh \calB}\sup_{\calB\in\setB_{N,\calS}}\E{\|\wh{\calB} - \calB\|_F}\nonumber\\
    & \!\!\!\!\geq \!\!\!\!& \frac{s}{2}\bigg(1- \frac{\frac{T}{2\gamma^2}\max_{j_1\neq j_2}\|\calB^{j_1} -\calB^{j_2} \|_F^2 + \log 2 }{\log n}    \bigg).
\end{eqnarray}

Next, we consider one construction for the sets of $\{\calB^j  \}_{j=1}^n\in\setB_{N,\calS}$ such that we can obtain a proper lower bound for $\min_{j_1\neq j_2}\|\calB^{j_1} -\calB^{j_2} \|_F$ and a proper upper bound for $\max_{j_1\neq j_2}\|\calB^{j_1} -\calB^{j_2} \|_F$.  Given the equivalence between any two tensors $\calB^{j_1} = [\mA_1^{j_1},\calS, \mA_2 ]$, $ \calB^{j_2} = [\mA_1^{j_2},\calS, \mA_2 ]$ and their canonical forms \cite{qin2024guaranteed}, the TT-SVD can be applied to transform these tensors into canonical formats $\calB^{j_1} = [\wt\mA_1^{j_1},\wt\calS, \wt\mA_2 ]$ and $ \calB^{j_2} = [\wt\mA_1^{j_2},\wt\calS, \wt\mA_2 ]$ with $R(\wt\calS)R^{\rm H}(\wt\calS) = \mId_{N}$ and $\wt\mA_2\wt\mA_2^{\rm H} = \mId_N$. Given any $\delta >0 $ and $N \geq C'$ by \cite[Lemma 5]{agarwal2012noisy}, we can construct a set of  $\{ \wt\mA_1^{1}, \dots, \wt\mA_1^{n_0} \}$  with cardinality $n_0\geq \frac{1}{4}e^{\frac{LPN}{128}}$ such that: (1) $\| \wt\mA_1^{j}\|_F = \delta$ holds for all $j = 1,\dots, n_0$, (2) $\| \wt\mA_1^{j_1} - \wt\mA_1^{j_2}\|_F \geq \delta$ for all $j_1,j_2\in[n_0], j_1\neq j_2$.

Due to $\|\calB^{j_1} - \calB^{j_2}\|_F = \| [ \wt\mA_1^{j_1} - \wt\mA_1^{j_2}, \wt\calS, \wt\mA_2 ] \|_F  = \|\wt\mA_1^{j_1} - \wt\mA_1^{j_2}\|_F$, we can further get
\begin{eqnarray}
    \label{two bounds in the minimax}
    &&\max_{j_1\neq j_2}\|\calB^{j_1} - \calB^{j_2}\|_F = \max_{j_1\neq j_2}\|\wt\mA_1^{j_1} - \wt\mA_1^{j_2}\|_F\leq 2\delta,\nonumber\\
    &&\min_{j_1\neq j_2}\|\calB^{j_1} - \calB^{j_2}\|_F = \min_{j_1\neq j_2}\|\wt\mA_1^{j_1} - \wt\mA_1^{j_2}\|_F\geq \delta.
\end{eqnarray}
Then we plug \eqref{two bounds in the minimax} into \eqref{lower bound of the minimax} and have
\begin{eqnarray}
    \label{lower bound of the minimax conlcusion}
    \inf_{\wh \calB}\sup_{\calB\in\setB_{N,\calS}}\E{\|\wh{\calB} - \calB\|_F}  & \!\!\!\! \geq  \!\!\!\!&  \frac{\delta}{2}\bigg(1- \frac{\frac{2T\delta^2}{\gamma^2} + \log 2 }{c_1LPN}    \bigg)\nonumber\\
    & \!\!\!\! \geq  \!\!\!\!& c_2 \sqrt{\frac{LPN}{T}}\gamma,
\end{eqnarray}
where the second inequality with a constant $c_2>0$ follows $\delta = c_3\sqrt{\frac{LPN}{m}}\gamma$ for a constant $c_3>0$.

Similarly, we can utilize the TT-SVD to transform $\calB^{j_1} = [\mA_1^{j_1},\calS, \mA_2 ]$ and $ \calB^{j_2} = [\mA_1^{j_2},\calS, \mA_2 ]$ into orthonormal formats $\calB^{j_1} = [\wt\mA_1,\wt\calS, \wt\mA_2^{j_1} ]$ and $ \calB^{j_2} = [\wt\mA_1,\wt\calS, \wt\mA_2^{j_2} ]$ with $L^{\rm H}(\wt\calS)L(\wt\calS) = \mId_{N}$ and $\wt\mA_1^{\rm H}\wt\mA_1 = \mId_N$. Then following the previous analysis, we can get
\begin{eqnarray}
    \label{lower bound of the minimax conlcusion1}
    \inf_{\wh \calB}\sup_{\calB\in\setB_{N,\calS}}\E{\|\wh{\calB} - \calB\|_F} = \Omega\bigg( \sqrt{\frac{UMN}{T}}\gamma \bigg).
\end{eqnarray}

Finally, we sum \eqref{lower bound of the minimax conlcusion}-\eqref{lower bound of the minimax conlcusion1} and calculate the average to derive
\begin{eqnarray}
    \label{lower bound of the minimax conlcusion ave}
    \inf_{\wh \calB}\sup_{\calB\in\setB_{N,\calS}}\E{\|\wh{\calB} - \calB\|_F} = \Omega\bigg(\sqrt{\frac{LPN + UMN}{2T}}\gamma \bigg).
\end{eqnarray}
\end{proof}

\section{Proof of \Cref{Upper bound of error difference multi IRS}}
\label{Proof of upper bound for multi IRS}

\begin{proof}
Utilizing the identical derivation as \eqref{wX and X star relationship_1}, we can directly acquire
\begin{eqnarray}
    \label{wX and X star relationship_1 multi}
    \frac{1}{T}\|\calX(\wh\calB_D  -  \calB^\star_D)\|_F^2 \leq \frac{2}{T}|\<  \calW, \calX(\wh\calB_D - \calB^\star_D) \>|.
\end{eqnarray}
First, according to Theorem 4 in the main paper, we have
\begin{eqnarray}
    \label{lower bound of difference multi}
    \frac{1}{T}\|\calX(\wh\calB_D  -  \calB^\star_D)\|_F^2 \geq (1-\delta_{UM})\|\wh\calB_D  -  \calB^\star_D\|_F^2.
\end{eqnarray}
Additionally, we can also rewrite $\frac{2}{T}|\<  \calW, \calX(\wh\calB_D - \calB^\star_D) \>|$ as follows:
{\small \begin{eqnarray}
    \label{upper bound of difference multi}
    &\!\!\!\!\!\!\!\!&\hspace{-0.5cm}\frac{2}{T}|\<  \calW, \calX(\wh\calB_D - \calB^\star_D) \>|\nonumber\\
     &\!\!\!\!\!\!\!\!&\hspace{-0.5cm} = \frac{2\|\wh\calB_D - \calB^\star_D\|_F}{T}\max_{\calH\in\setB_{2N}^D, \|\calH\|_F\leq 1} |\<\calH, \calW\times_3^2 \mX^*\>|\nonumber\\
    &\!\!\!\!\!\!\!\!&\hspace{-0.5cm}= \frac{2\|\wh\calB_D - \calB^\star_D\|_F}{T}\hspace{-1.1cm}\max_{  \|\mH_i\|\leq 1, i\in[D], \atop \|L(\wt\calS_i)\|\leq 1,i\in[D], \|\mH_0\|_F\leq 1}  \hspace{-1.3cm} |\<[\mH_D, \wt\calS_D,\dots, \mH_0], \calW\times_3^2 \mX^*\>|.
\end{eqnarray}}
The last line follows that $\|\calH\|_F^2 = \|\mH_0\|_F \leq 1 $ for a left-orthogonal TT form   in {Appendix}~A of the main paper.

Next, we can  construct an $\epsilon$-net $\{\mH_i^{(1)} , \dots, \mH_i^{(n_i)}  \}$ with the covering number $n_i\leq (\frac{4+\epsilon}{\epsilon})^{N_i N_{i+1}}, i\in[D]$ for the set of factors $\{\mH_i\in\C^{N_{i+1}\times N_i}:  \|\mH_i\|\leq 1   \}$ such that
\begin{eqnarray}
    \label{H1_proof1}
    \sup_{\mH_i: \|\mH_i\|\leq 1}\min_{p_i\leq n_i} \|\mH_i-\mH_i^{(p_i)}\|\leq \epsilon.
\end{eqnarray}

In addition, we can  construct an $\epsilon$-net $\{\mH_0^{(1)} , \dots, \mH_0^{(n_0)}  \}$ with the covering number $n_0\leq (\frac{2+\epsilon}{\epsilon})^{N_0 N_{1}}$ for the set of factors $\{\mH_0\in\C^{N_{1}\times N_0}:  \|\mH_0\|_F\leq 1   \}$ such that
\begin{eqnarray}
    \label{H0_proof1}
    \sup_{\mH_0: \|\mH_0\|_F\leq 1}\min_{p_0\leq n_0} \|\mH_0-\mH_0^{(p_0)}\|_F\leq \epsilon.
\end{eqnarray}

Therefore, we can construct an $\epsilon$-net $\{\calH^{(1)},\ldots,\calH^{(n_0\cdots n_D)}\}$ with covering number $n_0\cdots n_D \leq (\frac{4+\epsilon}{\epsilon})^{\sum_{d=0}^{D}N_d N_{d+1}}$ for any $\calH \in\setB_{2N,\{\wt\calS_i\}}^D$.

Denote by $A$ the value of \eqref{upper bound of difference multi}, i.e.,
\begin{eqnarray}
    \label{ProofOf<H,X>forSubGaussian_proof3 multi}
    &\!\!\!\!\hspace{-0.5cm}\!\!\!\!&[\wt\mH_D, \wt\calS_D,\dots,\wt\mH_1, \wt\calS_1, \wt\mH_0]\nonumber\\
    &\!\!\!\!\hspace{-0.5cm}=\!\!\!\!& \hspace{-0.7cm} \argmax_{  \|\mH_i\|\leq 1, i\in[D], \atop \|L(\wt\calS_i)\|\leq 1,i\in[D], \|\mH_0\|_F\leq 1} \hspace{-0.9cm} |\<[\mH_D, \wt\calS_D,\dots,\mH_1, \wt\calS_1, \mH_0], \calW\times_3^2 \mX^*\>|,
    \end{eqnarray}
    \begin{eqnarray}
    \label{ProofOf<H,X>forSubGaussian_proof4 multi}
    &\!\!\!\!\hspace{-0.5cm}\!\!\!\!&A:= \frac{1}{T}|\<[\wt\mH_D, \wt\calS_D,\dots,\wt\mH_1, \wt\calS_1, \wt\mH_0], \calW\times_3^2 \mX^*  \>|.
\end{eqnarray}
Using $\calI$ to denote the index set $[n_0]\times \cdots \times [n_D] $, then
according to the construction of the $\epsilon$-net, there exists $p=(p_0,\dots,p_D)\in\calI$ such that
\begin{eqnarray}
    \label{ProofOf<H,X>forSubGaussian_proof5 multi}
    \|\wt\mH_i - \mH_i^{(p_i)} \|\leq\epsilon, i\in[D] \ \text{and} \ \|\wt\mH_0 - \mH_0^{(p_0)} \|_F\leq\epsilon,
\end{eqnarray}
and taking $\epsilon=\frac{1}{2(D+1)}$ gives
{\small \begin{eqnarray}
    \label{ProofOf<H,X>forSubGaussian_proof6 multi}
    A&\!\!\!\!\leq\!\!\!\!&\frac{1}{T}|\<[\mH_D^{(p_D)}, \wt\calS_D,\dots, \wt\calS_1, \mH_0^{(p_0)}], \calW\times_3^2 \mX^*  \>|\nonumber\\
    &\!\!\!\!\!\!\!\!& + \frac{1}{T}|\<[\wt\mH_D, \wt\calS_D,\dots, \wt\calS_1, \wt\mH_0]\nonumber\\
    &\!\!\!\!\!\!\!\!& - [\mH_D^{(p_D)}, \wt\calS_D,\dots, \wt\calS_1, \mH_0^{(p_0)}], \calW\times_3^2 \mX^*  \>| \nonumber\\
    &\!\!\!\!=\!\!\!\!&\frac{1}{T}|\<[\mH_D^{(p_D)}, \wt\calS_D,\dots, \wt\calS_1, \mH_0^{(p_0)}], \calW\times_3^2 \mX^*  \>|\nonumber\\
     &\!\!\!\!\!\!\!\!&+ \frac{1}{T}\bigg|\sum_{i=0}^D\bigg\< [\mH_D^{(p_D)}, \wt\calS_D,\mH_{D-1}^{(p_{D-1})},\dots, \mH_{i+1}^{(p_{D_{i+1}})},\wt\calS_{i+1}, \nonumber\\
    &\!\!\!\!\!\!\!\!&\wt\mH_i-\mH_i^{(p_i)}, \wt\calS_i,\wt\mH_{i-1},\dots,\wt\calS_1,\wt\mH_0 ], \calW\times_3^2 \mX^* \bigg\> \bigg|\nonumber\\
    &\!\!\!\!\leq\!\!\!\!&\frac{1}{T}|\<[\mH_D^{(p_D)}, \wt\calS_D,\dots, \wt\calS_1, \mH_0^{(p_0)}], \calW\times_3^2 \mX^*  \>|\nonumber\\
    &\!\!\!\!\!\!\!\!& + (D+1) \epsilon A\nonumber\\
    &\!\!\!\!=\!\!\!\!&\frac{1}{T}|\<[\mH_D^{(p_D)}, \wt\calS_D,\dots, \wt\calS_1, \mH_0^{(p_0)}], \calW\times_3^2 \mX^*  \>| + \frac{A}{2},
\end{eqnarray}}
where the first equation follows the fact that $(l,k,u)$-th element of $[\wt\mH_D, \wt\calS_D,\dots, \wt\calS_1, \wt\mH_0] - [\mH_D^{(p_D)}, \wt\calS_D,\dots, \wt\calS_1, \mH_0^{(p_0)}]$ can be decomposed into $\sum_{i=0}^D\mH_D^{(p_D)}(l,:)\wt\calS_D(:,k,:)\mH_{D-1}^{(p_{D-1})} \\ \cdots\mH_{i+1}^{(p_{D_{i+1}})}\wt\calS_{i+1}(:,k,:)(\wt\mH_i-\mH_i^{(p_i)})\wt\calS_i(:,k,:)\wt\mH_{i-1}
\wt\calS_1(:,k,:)\wt\mH_0(:,u)$.

Based on the same analysis of \eqref{the tail function of fixed gaussian random variable 3}, we can obtain
 \begin{eqnarray}
    \label{the tail function of fixed gaussian random variable 3 multi IRS}
    &\!\!\!\!\!\!\!\!&\P{A \leq \frac{c_1\sqrt{(1+\delta_{UM})( \sum_{d=1}^{D-1} N_dN_{d+1})\log D}}{\sqrt{T}}\gamma }\nonumber\\
    &\!\!\!\! \geq\!\!\!\!& 1-2e^{-c_2( \sum_{d=1}^{D-1} N_dN_{d+1})\log D},
\end{eqnarray}
where $c_i,i=1,2$ are constants.

Combing \eqref{lower bound of difference multi}, we arrive at
{\small \begin{eqnarray*}
    \label{upper bound of error final multi IRS}
    \|\wh \calB_D - \calB^\star_D\|_F =  O\bigg(\frac{\gamma\sqrt{(1+\delta_{UM})(\sum_{d=1}^{D-1}N_d N_{d+1})\log D}}{(1-\delta_{UM})\sqrt{T}}\bigg).
\end{eqnarray*}}
\end{proof}

\section{Error Analysis of channel estimation in the Multi-Hop IRS-Assisted MIMO Systems with special structures}
\label{appendix:error analysis for discussion of structured case}

In this section, we present the error analysis of channel estimation in multi-hop IRS-assisted MIMO systems under three special structures, analogous to the single-hop case.

$(i)$ First, when $\mB_d, d=0,\dots D$ are rank-$r_d$ matrices, by incorporating the covering number of low-rank matrices into \Cref{Proof of upper bound for multi IRS}, we can derive
{\small \begin{eqnarray*}
    \label{upper bound of error final_multi IRS conclusion low-rank}
    \|\wh \calB_D - \calB^\star_D\|_F =  O\bigg(\frac{\gamma\sqrt{(1+\delta_{UM})(\sum_{d=0}^{D}(N_d + N_{d+1})r_d)\log D}}{(1-\delta_{UM})\sqrt{T}}\bigg),
\end{eqnarray*}}
where $\wh \calB_D,\calB^\star_D\in\{\calB_D = [\mB_D,\calS_D,\dots,\mB_1,\calS_1,\mB_0]\in\C^{LP\times K\times UM}:\mB_D\in\C^{LP\times N_D}, \mB_0\in\C^{N_1\times UM}, \mB_d\in\C^{N_{d+1}\times N_{d}},d\in[D-1] \  \text{are unknown with } \text{rank}(\mB_d) = r_d, d=0,\dots D, \ \text{and}  \ \calS_d\in\C^{N_d\times K\times N_d}, d\in[D] \ \text{are known} \}$.  Leveraging low-rank structures offers a potential avenue for improving error performance, as the recovery error may be reduced below eq. (14) in the main paper when the ranks $r_d, d=0,\dots, D$ are small.

$(ii)$ Second, when the channel matrices exhibit sparsity in the angular domain, we assume the factorized form $\mB_d = \mB_{d1} \mSigma_d \mB_{d2}, d=0,\dots D$, where $\mSigma_d\in\C^{F_d\times F_d}$ is a diagonal matrix with $\|\mSigma_d\|_0 = s_d$, and $\mB_{d1}\in\C^{N_{d+1}\times F_d}$, $\mB_{d2}\in\C^{F_d\times N_{d}}$. Following a similar approach as in \Cref{Proof of upper bound for IRS sparse}, we obtain the error bound:
{\small \begin{eqnarray*}
    \label{upper bound of error final_multi IRS conclusion sparse}
    \|\wh \calB_D - \calB^\star_D\|_F =  O\bigg(\frac{\gamma\sqrt{(1+\delta_{UM})(\sum_{d=0}^{D} s_d\log\big( \frac{F_d}{s_d} \big) )\log D}}{(1-\delta_{UM})\sqrt{T}}\bigg),
\end{eqnarray*}}
where $\wh \calB_D,\calB^\star_D\in \{\calB_D = [\mB_D,\calS_D,\dots,\mB_1,\calS_1,\mB_0]\in\C^{LP\times K\times UM}: \mB_d =   \mB_{d1} \mSigma_d \mB_{d2}, d=0,\dots D \ \text{where} \ \mSigma_d\in\C^{F_d\times F_d} \ \text{is  unknown}    \text{with} \ \|\mSigma_d\|_0 = s_d, \ \text{and} \ \mB_{d1}\in\C^{N_{d+1}\times F_d}, \mB_{d2}\in\C^{F_d\times N_{d}} \ \text{are known, and} \ \calS_d\in\C^{N_d\times K\times N_d}, d\in[D] \ \text{are known}  \}$. Note that the sparsity in the angular domain leads to a smaller recovery error compared to the non-sparse setting when $s_d \ll F_d$.

$(iii)$ Finally, when $\calS_i, i\in[D]$, are unknown, introducing the covering number of $\{\calS_i \}_{i=1}^{D}$ into the proof in  \Cref{Proof of upper bound for multi IRS}, we can arrive at
{\small \begin{eqnarray*}
    \label{upper bound of error final_multi IRS conclusion unknown S}
    &\!\!\!\!\!\!\!\!&\|\wh \calB_D - \calB^\star_D\|_F\nonumber\\
    &\!\!\!\!=\!\!\!\!& O\bigg(\frac{\gamma\sqrt{(1+\delta_{UM})(\sum_{d=0}^{D}N_d N_{d+1} + \sum_{d=1}^{D}K N_d )\log D}}{(1-\delta_{UM})\sqrt{T}}\bigg),
\end{eqnarray*}}
where $\wh \calB_D, \calB^\star_D\in \{\calB_D = [\mB_D,\calS_D,\dots,\mB_1,\calS_1,\mB_0]\in\C^{N_{D+1}\times K\times N_0}: \mB_d\in\C^{N_{d+1}\times N_{d}}, d=0,\dots,D, \calS_d\in\C^{N_d\times K\times N_d}, d\in[D] \  \text{are unknown} \}$. In this analysis, we assume that all IRSs are equipped with active elements. However, in practice, when adjustments are required for only a subset of IRSs, denoted as $\calS_i$ for $i \in \Omega$, where $\Omega \subseteq \{1, \dots, D\}$, the recovery error can be further reduced. Specifically, the term $\sum_{d=1}^D K N_d$ is replaced by $\sum_{i \in \Omega} K N_i$.

%


\end{document}